\documentclass[aps,pre,preprint,superscriptaddress,longbibliography]{revtex4-1}
\usepackage{amssymb}
\usepackage{tabularx}
\usepackage{setspace}
\usepackage[table]{xcolor}
\usepackage{xr}
\usepackage{color, colortbl}
\usepackage[section]{placeins}
\usepackage{caption}
\usepackage{subcaption}
\usepackage{amsmath}
\usepackage{graphicx}
\usepackage{makecell}
\usepackage{multirow}
\usepackage{tikz}
\usepackage{float}
\usepackage[utf8]{inputenc}
\usepackage[T1]{fontenc}
\usepackage{lmodern} 
\usepackage{diagbox}
\usepackage[a4paper, margin=1in]{geometry}
\usepackage{lipsum} 
\usepackage{rotating}
\usepackage{adjustbox}
\usepackage{tikz}
\usepackage{hyperref}
\hypersetup{
    colorlinks,
   citecolor=blue,
    filecolor=black,
    linkcolor=black,
    urlcolor=black,
    frenchlinks=true
}

\begin{document}

\title{A simplified digital twin of a pressure swing adsorption plant for air separation}

\author{Abhijit Dhamanekar}
\affiliation{Engineering Mechanics Unit, Jawaharlal Nehru Centre for Advanced Scientific Research, Jakkur, Bangalore, 560064, India}
\author{Ritwik Das}
\affiliation{Engineering Mechanics Unit, Jawaharlal Nehru Centre for Advanced Scientific Research, Jakkur, Bangalore, 560064, India}
\author{Santosh Ansumali}
\affiliation{Engineering Mechanics Unit, Jawaharlal Nehru Centre for Advanced Scientific Research, Jakkur, Bangalore, 560064, India}

\author{Raviraju Vysyaraju}
\affiliation{Department of Chemical and Materials Engineering, University of Alberta, Edmonton, T6G 2H5, Alberta, Canada}

\author{Arvind Rajendran}
\affiliation{Department of Chemical and Materials Engineering, University of Alberta, Edmonton, T6G 2H5, Alberta, Canada}

\author{Diwakar S. V.}
\email{diwakar@jncasr.ac.in}
\affiliation{Engineering Mechanics Unit, Jawaharlal Nehru Centre for Advanced Scientific Research, Jakkur, Bangalore, 560064, India}

\begin{abstract}

The pressure swing adsorption (PSA) process is one of the widely utilized techniques for air separation. Operating on the Skarstrom cycle, the porous adsorbent columns of a PSA system alternate between adsorption and desorption phases to selectively enrich the desired component in a gas mixture. The current work presents a robust and generalizable digital twin CFD model of a PSA system that can significantly help in design and device characterization. Using an axisymmetric representation, the digital twin accurately mimics all the key components of an air separation plant, including the air reservoir, adsorbent columns, product buffer tank, pressure regulator, solenoidal valves, and mesh filters. The model simulates the flow and adsorption processes in the system by solving the conservation equations for mass, momentum, energy, and species, along with the equation for adsorption kinetics. The cyclic operation of the PSA plants, typically controlled by solenoid valves, is emulated by dynamically modifying the boundary conditions of different subdomains. Such an integrated approach is shown here to closely replicate the performance of an in-house PSA pilot setup producing oxygen in terms of purity and pressure transience. Also, both the numerical and the experimental results yield an optimum performance for the same process parameters, such as pressurization time (26 s), purge time (2 s), and equalization time (4 s). The proposed numerical model is versatile and can be adapted to various industrial applications of PSA technology, such as hydrogen purification and carbon capture. Thus, it offers a cost-effective tool for designing and optimizing PSA systems.

\end{abstract}



\maketitle

\section{Introduction}
\label{sec_intro} 
Air constituents such as nitrogen, oxygen, and argon are indispensable in both medical and industrial applications. The separation of these constituents is commonly achieved through techniques such as cryogenic distillation \cite{Miller}, membrane separation \cite{Richard_W_Baker}, and pressure swing adsorption (PSA) \cite{Ackley2019, ARVIND2002419, Hejazi}. Among these, PSA is widely employed for commercial purposes, including the separation of $\mathrm{O_{2}}$ and $\mathrm{N_{2}}$ from air \cite{SIRCAR, SIRCAR1989, Ackley2019}, hydrogen production through natural gas reforming \cite{SIRCAR}, and $\mathrm{CO_{2}}$ capture and sequestration \cite{RIBOLDI20172390}. The PSA technique operates on the principle of selective adsorption, wherein specific gas molecules in a mixture, when pressurized, are better adsorbed in porous materials. For example, nitrogen molecules are preferentially adsorbed onto materials like zeolites due to their higher quadrupole moment compared to oxygen and argon \cite{peter2005adsorption}. This selective adsorption enables the production of high purity of $\mathrm{O_2}$ and $\mathrm{Ar}$ gas mixtures. A conventional PSA process typically employs a two-column configuration that follows a modified Skarstrom cycle \cite{MENDES2001173}, wherein the columns alternate between the adsorption and desorption phases, ensuring a continuous supply of the desired product gas. Additional intervening steps, such as purge and pressure equalization, can enhance product purity and recovery. Interestingly, PSA systems offer several advantages: 1) due to their shorter cycle time, they quickly arrive at the desired purity; 2) they involve lower maintenance costs compared to membrane separation; and 3) their flexibility and compactness make them suitable for in-situ gas generation. Although the output purity of PSA plants is lower than that of cryogenic distillation systems, their simplicity and cost-effectiveness make them the preferred choice for small-scale applications \cite{Ackley2019, MENDES2001173}.

Decades of research and development have notably advanced PSA technology, the relevance of which became globally evident during the COVID-19 pandemic \cite{GHES01589}. Nevertheless, there is an enormous potential for improvement in various aspects such as a) the development of new adsorbent materials with higher extract production capacity, higher selectivity, better mechanical strength \& stability, and low cost \cite{Baksh, Jayaraman, Nandi}, b) optimization of process cycle in terms of their sequence and timings \cite{Mahdi}, and c) making better flow distributors for reducing pressure drop \cite{Moran, Bozzano}. In this regard, much success can be harnessed through numerical modelling, which can help save significant cost and effort involved in building and testing new experimental prototypes. Detailed perspectives of various numerical works relating to PSA systems can be obtained from the reviews of Zhang et al. \cite{pr10050812}. Incidentally, most of these works focus on 1D models that integrate mass, momentum, energy, and species conservation equations over time to emulate the transient response of the PSA plants in a simplified way. The process of integration is often aided by supplementary equations that model adsorption via a simple Linear Driving Force (LDF) approach. With an axially dispersed plug flow consideration, these models effectively predict the transient variation of mass, velocity, temperature, and composition fields within the 1D domain \cite{Jee} and help superficially validate lab-scale experimental setups \cite{Ferreira}. Unfortunately, these simple models do not account for the lateral variation in field variables. For example, the heat loss through the column walls, particularly relevant for small systems, is often ill-estimated. The models also do not account for the intricate features of flow distributors, valves, and other components that bring notable performance improvements to commercial plants. Thus, developing 3D or 2D axisymmetric models is essential for predicting the behaviour of commercial PSA plants. However, such works are very few in the published literature. Gautier et al.~\cite{GAUTIER2018314} have developed a 3D CFD model to simulate the PSA process for $\mathrm{CO_{2}}$/$\mathrm{CH_{4}}$ gas separation and also performed a sensitivity analysis to determine the influence of different operating parameters on the process performance. Recently, Ramos et al. \cite{FABIANRAMOS} developed a 2D formulation wherein the transport phenomena and the adsorption processes have been modelled in the Ansys-Fluent package using User Defined Functions (UDFs).

The current work extends the above framework to develop a simplified digital twin of a commercial-type PSA plant. The primary goal is to model the entire PSA system and not just the adsorbent columns, as typically done in many published literature. Starting from the buffer tank of the compressor, the present numerical model includes all the essential components of the plant, such as adsorbent columns, product storage tank, solenoid valves, pressure regulators, and mesh filters. The detailed mathematical formulation incorporated here conserves mass, momentum, energy, and species in all these components via a 2D axisymmetric representation in Ansys-Fluent. Similar to the work of Ramos et al.\cite{FABIANRAMOS}, the adsorption kinetics are integrated with the conservation equations through User Defined Functions (UDFs). The most notable feature of the present digital twin is the realization of solenoid valves' operation by modifying various sub-domain boundary conditions, i.e., the intervening boundaries between the columns are appropriately switched between wall and interface boundary conditions. Consequently, the digital twin accurately predicts the transient buildup of pressure and output purity in the buffer tank for different timings of the pressurization, purge, and equalization steps. Note that the present simulations have been carried out in the context of oxygen separation from air using 13X zeolites. The results of the simulations have been validated against the experiments carried out on our in-house PSA pilot plant, which is capable of producing around 20 slpm of oxygen product at 93\% purity. The details of this experimental plant are presented in the next section, followed by information on the digital twin and the comparison of results. Though the simulation framework presented here has been tested exclusively for $\rm O_2$ concentration, it can be easily extended to other scenarios, thus providing a valuable design tool for any system modification and scale-up.  

\section{Experimental setup}
\begin{figure}[!ht]
    \includegraphics[width = 6.5cm]{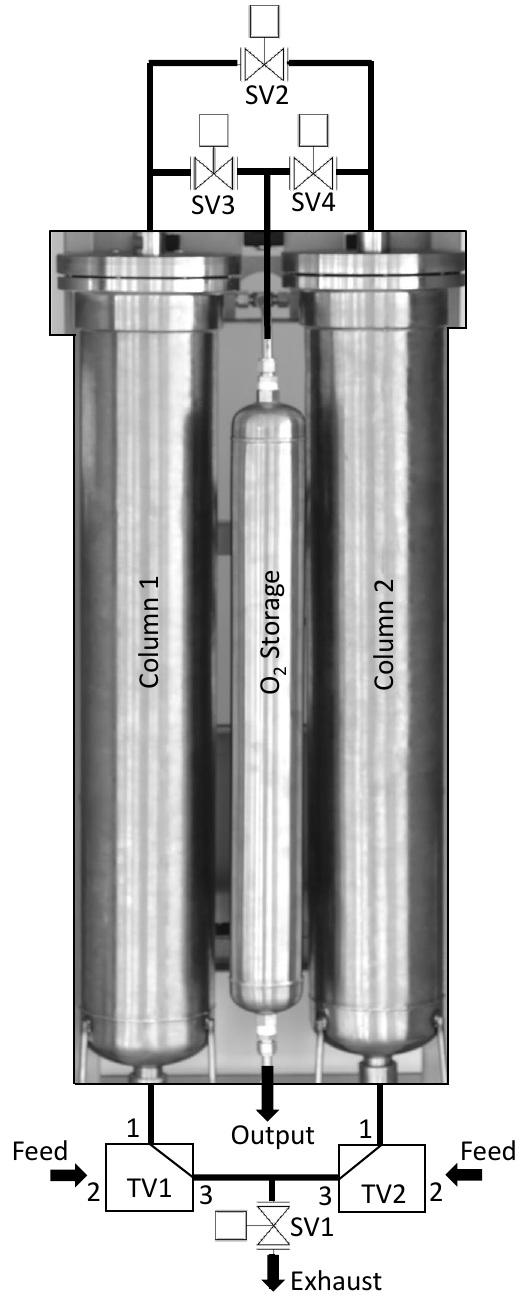}
    \centering
    \caption{The schematic of the experimental PSA pilot plant}
    \label{fig:schematic}
\end{figure}

The current experimental setup incorporates a modified version of the Skarstrom cycle, wherein two zeolite columns are alternately pressurized and depressurized to generate a sustained output of oxygen-enriched gas mixture. The schematic of the setup involving the two adsorbent columns, a product storage tank, and different solenoid valves is shown in Fig. \ref{fig:schematic}. A total of six solenoid valves from ASCO Valve Inc have been used, of which two are of the 3/2 type (TV1 and TV2), and the remaining four are of the bidirectional 2/2 type (SV1, SV2, SV3, and SV4). By default, ports 1 and 3 of the 3/2 valve are normally open, and ports 1 and 2 connect when the valve is energized. All the 2/2 valves are in a normally closed configuration. 

\begin{figure}[!ht]
    \includegraphics[width = 14cm]{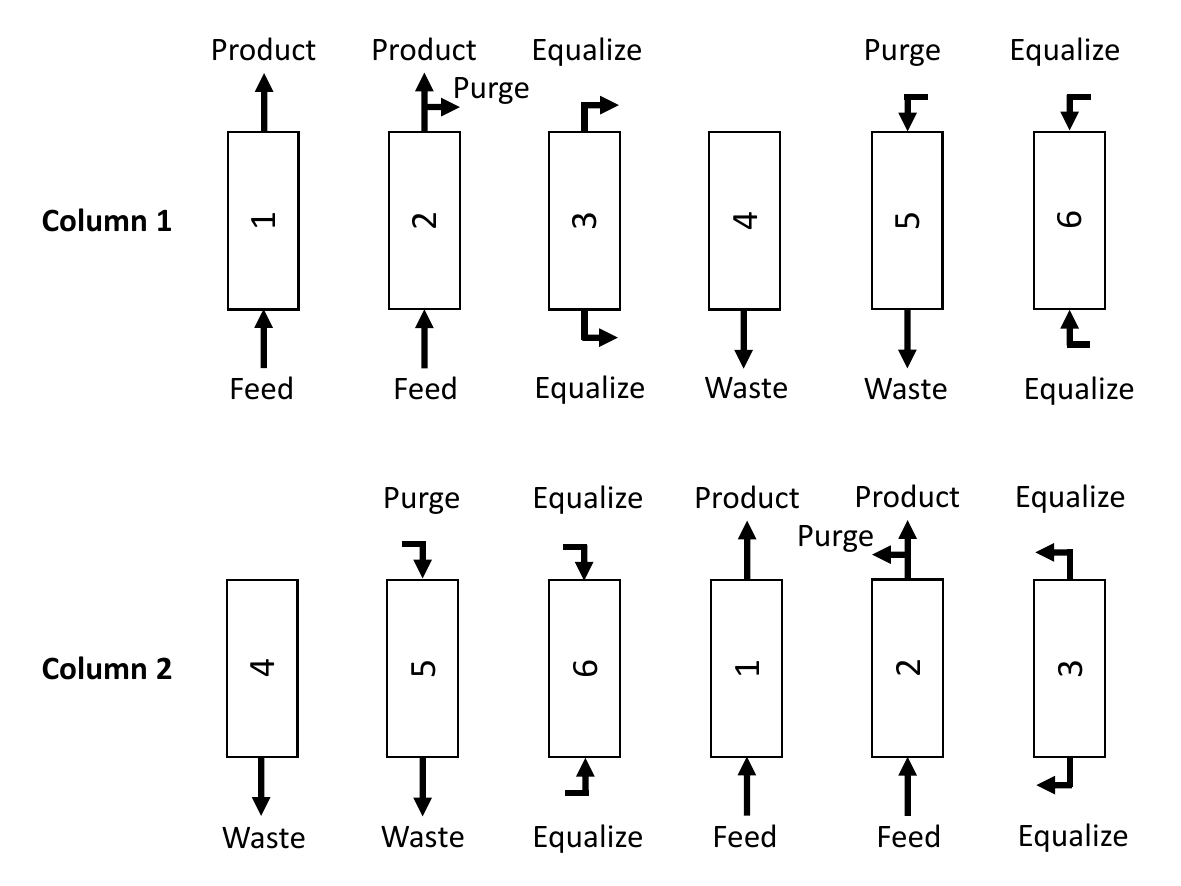}
    \centering
    \caption{Modified Skarstrom cycle}
    \label{fig:exp_cycle}
\end{figure}

\begin{table}
\renewcommand{\arraystretch}{1.2}
\centering
\begin{tabular}{|l|c|c|c|c|c|c|c|c|}
  \hline
  \multirow{2}{*}{\diagbox[width=4cm]{\textbf{Step}}{\textbf{Valve ID}}} 
  & \multicolumn{2}{c}{{\textbf{TV1}}} & \multicolumn{2}{|c|}{{\textbf{TV2}}} & \multirow{2}{*}{\textbf{SV1}} & \multirow{2}{*}{\textbf{SV2}} & \multirow{2}{*}{\textbf{SV3}} & \multirow{2}{*}{\textbf{SV4}} \\\cline{2-5} 
    & 1-2 & 1-3 & 1-2 & 1-3 & & & & \\
   \hline \hline
   \textbf{Pressurization-1} & \textbf{1} & 0 & 0 & \textbf{1} & \textbf{1} & 0 & \textbf{1} & 0 \\
   \hline
   \textbf{Purge-1} & \textbf{1} & 0 & 0 & \textbf{1} & \textbf{1} & \textbf{1} & \textbf{1} & 0 \\
   \hline
   \textbf{Equalization-1 }& 0 & \textbf{1} & 0 & \textbf{1} & 0 & \textbf{1} & 0 & 0 \\
   \hline
   \textbf{Pressurization-2} & 0 & \textbf{1} & \textbf{1} & 0 & \textbf{1} & 0 & 0 & \textbf{1} \\
   \hline
   \textbf{Purge-2} & 0 & \textbf{1} & \textbf{1} & 0 & \textbf{1} & \textbf{1} & 0 & \textbf{1} \\
   \hline
   \textbf{Equalization-2} & 0 & \textbf{1} & 0 & \textbf{1} & 0 & \textbf{1} & 0 & 0 \\
   \hline
 \end{tabular}
\caption{\label{tab:exp_valve_config} Valve Sequencing for different steps of the Modified Skarstrom cycle. \textbf{1}-Open, 0-Close.}
\end{table}

The various steps of the present Skarstrom cycle implementing the pressurization, purge, equalization, and depressurization processes are shown in Fig.~\ref{fig:exp_cycle}. Here, the top row indicates the six steps sequentially executed in Column 1, while the bottom row illustrates the steps that are simultaneously performed in Column 2. Table~\ref{tab:exp_valve_config} presents the valve sequencing implementation that effectuates the above steps. In the first step, Column-1 is pressurized with high-pressure feed air by energizing valve TV1 (port 1-2). Valve SV3 is opened for high-purity oxygen delivery, and valves SV2 and SV4 are kept closed. Meanwhile, Column-2 is depressurized by turning off valve TV2 (port 1-3) and opening valve SV1. In the second step, all the valves are maintained in the same condition except SV2, which is now opened to allow for the purging of Column-2 with an enriched oxygen mixture from Column-1. In the third step, the pressures in the two columns are equalized by keeping all the valves, except SV2, closed. There is no input of feed air to the system and no extraction of oxygen-rich media from the zeolite columns. The high-pressure gas stored in the buffer tank compensates for the supply intermittency during this step. Note that the equalization is performed at both ends of the columns in the present implementation of the cycle. The fourth step is the opposite of the first step. Column-1 undergoes depressurization, while Column-2 is pressurized. Valves TV2 (port 1-2), SV1, and SV4 are energized, and all the other valves are kept closed. In the fifth step, valve SV2 is once again opened along with the above valves to purge Column-1. The last (equalization) step is precisely the same as the third step, and at the end of this step, the system's state becomes amenable to the cycle's repetition. 

\begin{table}[!h]
\renewcommand{\arraystretch}{1.1}
\begin{tabular}{|l|c|}
 \hline
  
  \multicolumn{2}{|c|}{\centering \textbf{Adsorbent cylinder}} \\
  \hline \hline
   Cylinder Material  & SS304  \\
   \hline
  Effective Length, $L_{zeo}$  & 930 mm\\
  \hline
  Inner Diameter, $D_{zeo}$ & 162.8 mm \\
  \hline
  Cylinder Thickness, $th_{zeo}$ & 2.77 mm \\
  \hline
  Filter Mesh & 100 $\mu m$\\
  \hline \hline
  \multicolumn{2}{|c|}{\centering \textbf{$O_2$ Cylinder}} \\
  \hline \hline
  Total Length, $L_{O_2}$  & 568 mm\\
  \hline
  Inner Diameter, $D_{O_2}$ & 66.9 mm \\
  \hline
  Cylinder Thickness, $th_{O_2}$ & 3.05 mm \\
  \hline \hline
  \multicolumn{2}{|c|}{\centering \textbf{Adsorbent}} \\
  \hline \hline
  Commercial Name  & UOP-HP 13X \\
  \hline
   Mass of adsorbent in each cylinder & 12.9 Kg \\
  \hline
  Mean particle diameter & 1.5 mm \\
  \hline \hline
  \multicolumn{2}{|c|}{\centering \textbf{Others}} \\
   \hline \hline
  Reported $C_V$ of the Solenoid Valves  & 1.5 \\
  \hline
  Compressor Size & 5HP (oil-free) \\
 \hline
 Air Reservoir Size & 200 litres \\
 \hline
  Free Air Delivery (FAD) & $\approx$ 15 CFM \\
  \hline
  Air Dryer &  Dessicant-based \\
  \hline
 Dryer Purge Loss & $\leq$ 15\%  \\
    \hline 
\end{tabular}
\centering
\caption{\label{tab:PSA_exp_det} Details of the experimental setup}
\end{table}

Table~\ref{tab:PSA_exp_det} shows the relevant details of the current experimental setup. As evident in Fig.~\ref{fig:schematic}, the adsorbent columns have a 2:1 ellipsoidal dish end at the bottom and a flange cover at the top. A spring-piston arrangement was utilized in both columns to arrest the zeolite particles' movement and subsequent fluidization. Both ends of the oxygen storage tank were attached with a 2:1 ellipsoidal dish end. The columns and the solenoidal valves were interconnected via 1/2-inch diameter stainless steel pipes. The regulated feed air supply for the PSA system was provided by a 5HP oil-free air compressor with a Free Air Delivery (FAD) of approximately 15 cfm. A desiccant-based air dryer was used to remove moisture from the feed air to the extent of -40$^{\circ}$C Pressure Dew Point (PDP). The ON-OFF sequencing of the solenoidal valves was effectuated through a microcontroller-based timer. The pressures in the individual adsorbent columns were measured using Honeywell PX2 series pressure transducers. A Honeywell AWM700 series flow sensor and a Honeywell OOM202 series oxygen sensor measured the output flow rate and oxygen purity, respectively. The maximum feed air pressure was regulated at 7 bar, and the outlet flow rate of oxygen was maintained constant using a combination of a pressure regulator and a needle valve. 

A series of experimental trials have been performed on the above prototype plant to obtain the optimum timing for each step of the Skarstrom cycle, i.e., the pressurization (t$\mathrm{_{pr}}$), purge (t$\mathrm{_{pu}}$), and equalization (t$\mathrm{_{eq}}$) times. Before understanding the results of these trials, we describe the digital twin of the above PSA system in the ensuing section.

\section{Description of the PSA Digital twin}

\begin{figure}[]
    \includegraphics[width = 10.5cm]{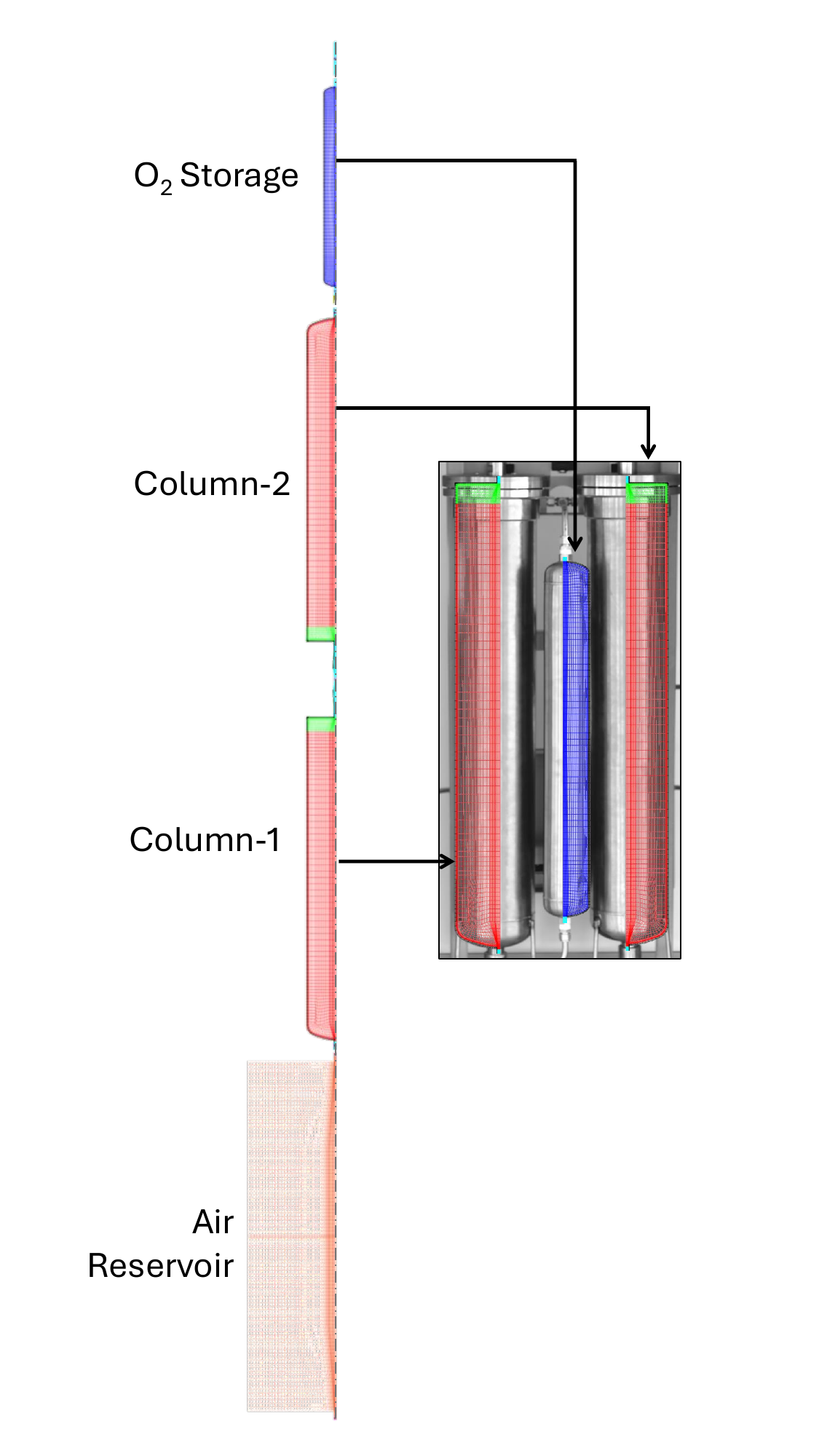}
    \centering
    \caption{The meshed geometry of digital twin model}
    \label{fig:mesh_dtwin}
\end{figure}

\begin{figure}[]
    \includegraphics[width = 10.5cm]{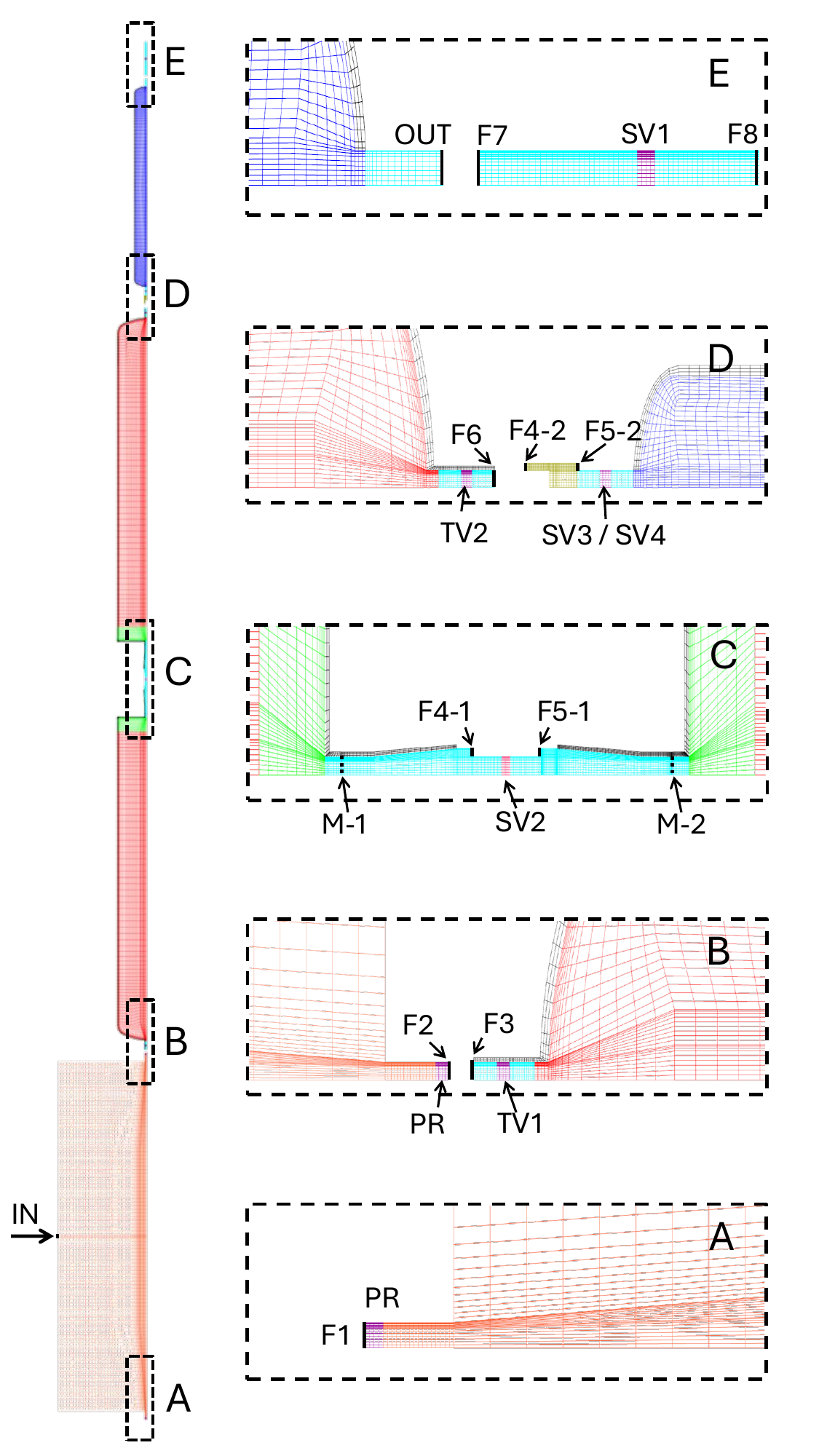}
    \centering
    \caption{Magnified sub-sections of the digital twin indicating pressure regulator (PR), solenoid valves (TV1, TV2, SV1, SV2, SV3, SV4), mesh filter (M1, M2), and different faces (IN, F1 to F8, OUT).}
    \label{fig:subsec_dtwin}
\end{figure}

The present digital twin of the PSA plant involves numerical models that accurately represent the plant's various components, such as air reservoir, adsorbent columns, oxygen storage tank, pressure regulator, solenoid valves, filters, and piping. Here, we employ an axisymmetric description of these components to substantially benefit from the lower computational costs in terms of memory usage and computation time. Unfortunately, not all geometries exhibit perfect rotational symmetry. Also, specific flow phenomena like the secondary flows and three-dimensional instabilities might be inadequately captured under an axisymmetric consideration. In particular, challenges arise from the interconnectivity of adsorbent columns, feed air supply from the reservoir, product output to the storage tank, T-joints, and solenoid valves. The current work overcomes all these difficulties by carefully adjusting the domain geometries and the respective boundaries while solving the various conservation equations using the Ansys-Fluent (v2020-R1) commercial package. Consequently, the current axisymmetric digital twin efficiently emulates the PSA-based gas separation process, illustrated here through a systematic comparison with the experimental data of the pilot plant described in the previous section. To begin with, we now present an overall description of the digital twin in the following subsection.

\subsection{The Overall System}

Figure~\ref{fig:mesh_dtwin} shows the meshed geometry of the digital twin model along with the real-world association of major components like the adsorbent columns and the oxygen storage tank. Magnified views of the different sub-sections of the model are shown in Fig.~\ref{fig:subsec_dtwin}. Various sub-components like pressure regulators (PR), solenoid valves (TV1 \& TV2 and SV1 to SV4), mesh filters (M1 and M2), and face interfaces (F1 to F8) have also been marked in Fig.~\ref{fig:subsec_dtwin}. Here, the pre-processing tasks, such as geometry creation and mesh generation, have been performed using the ICEM-CFD package. 

The overall model involves an air reservoir with its inlet marked as `IN' in Fig.~\ref{fig:subsec_dtwin}. Air is fed into the reservoir at a constant mass flow rate corresponding to the compressor's corrected FAD, which accounts for the elevated temperature of the air emanating from the compressor. In contrast to the experimental configuration wherein a single pressure regulator is used to supply air to the adsorbent columns, two pressure regulators (one at each end of the reservoir) have been employed here to accommodate its functionality in the axisymmetric configuration. These regulators have been marked as `PR' in the sub-figures A and B of Fig.~\ref{fig:subsec_dtwin}. The desiccant-based dryer has not been explicitly included in the model, though its effects, including the purge loss and pressure drop, have been suitably included. Accordingly, the air mass flow rate from the compressor to the reservoir is reduced by 15\% to compensate for the purge loss in the dryer. At the same time, the pressure drop in the dryer system is added to the pressure regulator zone. Modeled as a porous media, the zone restricts the downstream pressure of the regulator when reservoir pressure exceeds the set limit. 

Figure~\ref{fig:subsec_dtwin} also shows purple-coloured mesh regions that correspond to the six solenoid values utilized in the pilot PSA plant. The ON-OFF operation of these valves has been implemented by modifying the boundary condition of the face adjoining each valve. In other words, any valve is opened by specifying the adjoining face as an interface (ON) and is closed when it is specified as a wall boundary (OFF). In the present configuration, faces F3 and F7 control valves TV1 and TV2, respectively, faces F4-2 and F5-2 control valves SV3/SV4, and faces F7 and F8 control valve SV1. Only in the case of SV2, the valve region is altered between fluid and solid interiors to switch between ON and OFF states. Interestingly, the structure and the operating mechanisms of the solenoid valves contribute to a pressure drop even in their fully opened state. This has been modeled presently by resorting once again to the porous zone approach, wherein the pressure drop in the purple-coloured mesh regions is calculated as per the flow coefficients ($\rm K_V$) specified by the valve manufacturer. 

Apart from the six solenoid values and the pressure regulator, the sub-domains of Fig.~\ref{fig:subsec_dtwin} also show twelve different faces that help replicate the experimental cycle in the digital twin. Of these twelve, ten faces (except IN and OUT) form interface pairs that help connect different parts of the geometry. In this regard, conformal mesh mapping is performed using periodic boundary conditions (interface) in Ansys-Fluent. The face combinations that are selectively interfaced are F1 \& F6, F2 \& F3, F3 \& F8, F4-1 \& F4-2, F5-1 \& F5-2, and F6 \& F7. The periodic boundary condition allows for one-to-one mapping between the face pairs that are geometrically and mesh-wise identical. The translational type of periodic boundary conditions utilized here require the two constituent faces to be parallel to each other and their meshes identical so that a single translation transformation can be used by providing an offset. Note that the face pairs mentioned above would be of wall boundary type by default, and they get interfaced based on the step involved in the modified Skarstorm cycle. 

In the experimental pilot plant shown in Fig.~\ref{fig:schematic}, the gas line connections between the adsorbent columns and the product storage tank are not amenable for their direct representation in the axisymmetric model owing to their physical configuration. To resolve this issue, a T-joint connection involving faces F4-2 and F5-2, as shown in the sub-domain D of Fig.~\ref{fig:subsec_dtwin}, has been utilized. Here, face F4-2 will interface with face F4-1 of sub-domain C (Fig.~\ref{fig:subsec_dtwin}) when the oxygen-enriched gas needs to be delivered from Column-1 to the $\rm O_2$ storage tank. Similarly, faces F5-2 and F5-1 will form an interface pair when the delivery occurs from Column-2 to the $\rm O_2$ buffer tank. Note that the zeolite columns (sub-domain C of Fig.~\ref{fig:subsec_dtwin}) have two pathways for letting out the product gas: one through the solenoid valve SV2 that opens during the purging and equalization steps and the second through the faces F4-1 and F5-1 that connect Columns 1 and 2 with the oxygen storage tank during the product delivery step. In order to maintain consistency with the experimental setup, the cross-sectional areas of the faces F4 and F5 have been specified to be the same as those of the 1/2" piping. Also, the length of the pipes utilized in the model measures the same as those used in the pilot plant. 

In addition to the above modifications of the system geometry, an exhaust pipe, controlled by faces F7 and F8 (subdomain E), has been added to the model to ensure a proper discharge of the desorbed and purged gases. During the discharge from the first column, face F3 is interfaced with face F8, and face F7 is specified as an outflow boundary. Similarly, in the case of the second column, faces F6 and F7 are interfaced, and face F8 becomes the outflow boundary. Note that the above exhaust pipe is essential in the current model to account for the passage of discharging gas through two solenoid valves (TV1/TV2 and SV1) in the experimental setup. Incidentally, the feed gas passes through a single valve before entering the zeolite column. The face OUT in subdomain E is specified as a mass outflow condition wherein the outflow rate is set as per the experimental trials.

With different entities of the digital twin defined above, we now list the sequence of steps performed in the numerical simulations to show how they exactly mimic the experimental prototype. 

\begin{figure}[]
    \includegraphics[angle=90,width = 11.5cm]{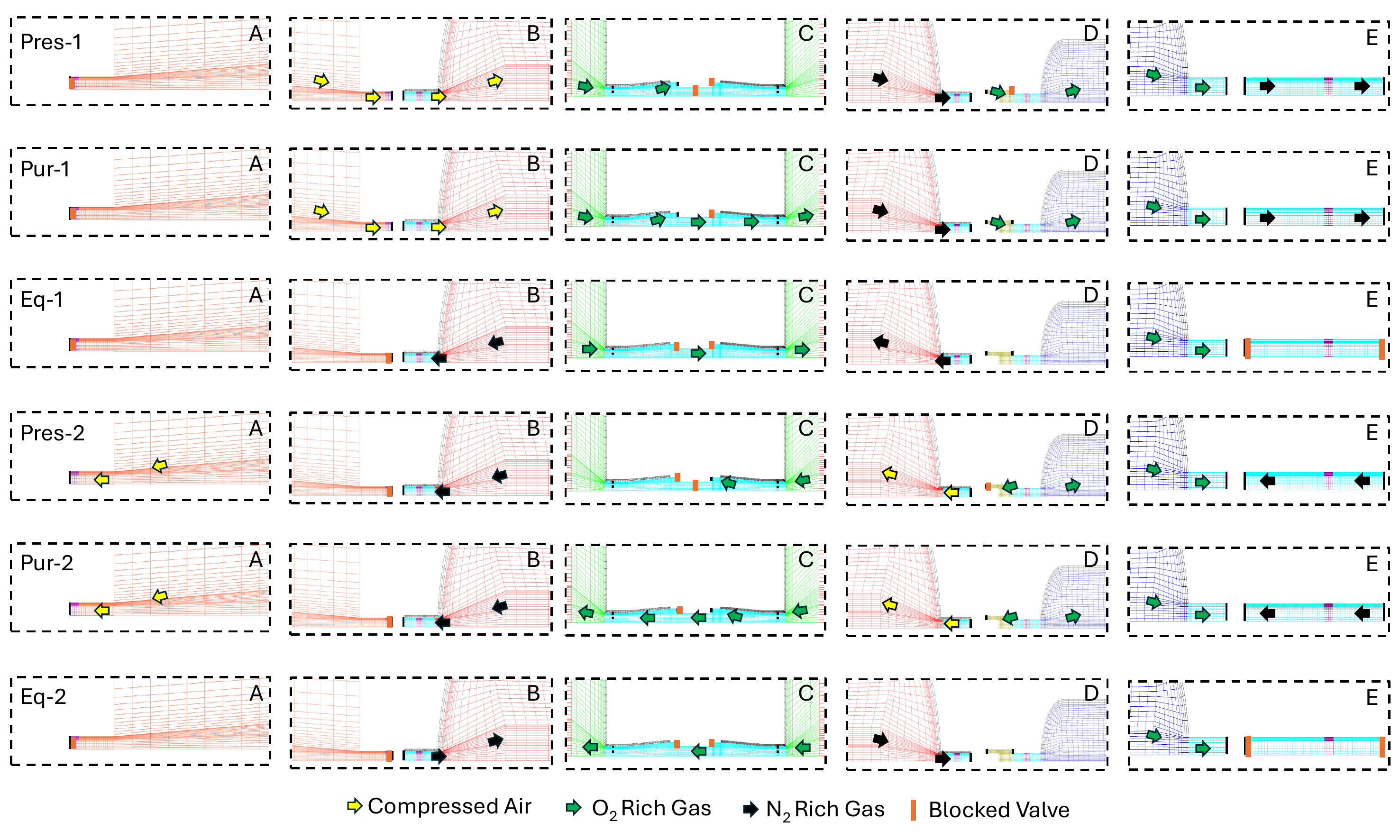}
    \centering
    \caption{Digital valve sequence and the resulting flow paths for different steps of the modified Skarstorm cycle}
    \label{fig:cycle_arrow_dtwin}
\end{figure}

\begin{itemize}

\item \textbf{Pressurization 1}: In this step, the compressed air is supplied from the air reservoir to zeolite Column-1 through the pressure regulator and valve TV1. In this regard, face F2 is interfaced with face F3 and face F1 is set as a wall. Face F4-1 is interfaced with face F4-2 to deliver the output gas to the storage tank. Simultaneously, depressurization occurs in cylinder 2. For this sake, faces F6 and F7 are interfaced, and a pressure outlet boundary condition is set at face F8. This ensures that the exhaust passes through two solenoid valves, viz. TV2 and SV1. Faces F5-1 and F5-2 are set as walls, and the domain SV2 is set as solid interior. 

\item \textbf{Purge 1}: The face configurations remain the same as in the previous step, except for SV2, which is now changed to a porous zone mimicking flow through the valve. This allows a small fraction of enriched product from Column-1 to purge Column-2.

\item \textbf{Equalization 1}: In order to equalize the pressure between the two columns, all the face interfaces mentioned in the first step are disconnected and are set as walls. Face F3 is now interfaced with face F6 to achieve bottom-bottom equalization. At the same time, valve SV2 is maintained as a fluid interior to enable top-top equalization. Since faces F1 and F2 are walls presently, there is no supply of feed air to the columns, and similarly, there is no enriched gas transfer to the storage tank since F4-1, F4-2, F5-1, and F5-2 have all been changed to walls. 

\item \textbf{Pressurization 2}: This step is the opposite of Pressurization 1 step. The compressed air is supplied to Column-2 through a pressure regulator and solenoid valve TV2. Face F1 is interfaced with face F6, and face F2 is set as a wall. Face F5-1 is interfaced with Face F5-2 to deliver enriched gas from Column-2 to the storage tank. In order to depressurize Column-1, face F3 is interfaced with face F8, and face F7 is changed from wall boundary condition to a pressure outlet. The flow now exits through two solenoid valves, TV1 and SV1. Faces F4-1 and F4-2 are kept as walls, and the domain SV2 is changed from a fluid interior to a solid interior.

\item \textbf{Purge 2}: In this purge step, face SV2 is again changed from a solid to a fluid interior to allow for the purging of Column-1. All the other face configurations remain the same as above. 

\item \textbf{Equalization 2}: The configuration of faces and valve domains are the same as those in the Equalization 1 step.
\end{itemize}

Figure~\ref{fig:cycle_arrow_dtwin} pictorially shows the above sequence modifications in the boundary conditions implemented in the present digital twin. The same has also been listed in Table~\ref{tab:valve config} for the reader's convenience. 

\begin{table}
\renewcommand{\arraystretch}{1.2}
\centering
\resizebox{\textwidth}{!}{%
\begin{tabular}{|l|c|c|c|c|c|c|c|c|c|c|c|}
  
  \hline
   \diagbox[width=4cm]{\textbf{Step}}{\textbf{Face ID}} & \textbf{F1} & \textbf{F2} & \textbf{F3} & \textbf{F4-1} & \textbf{F4-2} & \textbf{F5-1} & \textbf{F5-2} & \textbf{F6} & \textbf{F7}  & \textbf{F8} & \textbf{SV2}\\
   \hline \hline
   \textbf{Pressurization-1} & W & F3 &  F2 &  F4-2 &  F4-1 &  W & W & F7 & F6 & OT &  W \\
   \hline
    \textbf{Purge-1} & W & F3 &  F2 &  F4-2 &  F4-1 &  W & W & F7 & F6 & OT &  IT \\
   \hline
   \textbf{Equalization-1} & W & W &  F6 &  W & W & W & W & F3 & W  & OT & IT\\
   \hline
   \textbf{Pressurization-2} & F6 & W & F8 &  W & W & F5-2 & F5-1 & F1 & OT & F3 & W\\
   \hline
   \textbf{Purge-2} & F6 & W & F8 &  W & W & F5-2 & F5-1 & F1 & OT & F3 & IT\\
   \hline
   \textbf{Equalization-2} & W & W &  F6 &  W & W & W & W & F3 & OT & W & IT\\
   \hline
 \end{tabular}}
\caption{\label{tab:valve config}Valve Sequencing in terms of boundary conditions (W - Wall, OT - Outlet, IT - Interior)}
\end{table}

In addition to the above boundary manipulations, the present CFD model estimates the flow process through the zeolite particles via a porous zone approach. The region filled with the zeolites is shown in red colour in Fig.~\ref{fig:mesh_dtwin}. Since the particles are kept under compression using a spring-piston arrangement to prevent fluidization, a small clear fluid (non-porous) region, shown in green, is present within the columns. The pressure drop in the porous zone is accordingly calculated using a packed bed model involving Ergun's equation. The adsorption/desorption processes in the columns are emulated here by incorporating appropriate source terms in the continuity, energy, and species transport equations. The adsorption kinetics are modelled through the Linear Driving Force (LDF) approach. 

One can also observe that both the experimental adsorbent columns have been fitted with a mesh filter at the top to prevent the zeolite particles from escaping. For a proper estimation of the overall pressure drop, it is imperative to include the effects of these filters (marked as `M-1' and `M-2' in the sub-figure C of Fig.~\ref{fig:subsec_dtwin}) in the digital twin model. Correspondingly, a porous media approach is again invoked wherein the pressure drop across this metallic woven screen is calculated based on the flow rate. 

The influence of all the porous zones mentioned till now would manifest in the digital twin model through the momentum conservation equation. While these details will be discussed at length in sub-section 3.2.2, we now begin the process of reviewing all the conservation equations applicable throughout the domain.

\subsection{The conservation equations}

A realistic emulation of the adsorption phenomena in a PSA plant involves transient solutions of mass, momentum, energy, and species conservation equations, along with the modelling of the gas adsorption process. The present axisymmetric consideration amalgamates different porous and non-porous zones to arrive at a common framework of conservation equations. Consequently, special source terms are defined for the porous zones that mimic specific components like the adsorbent columns, pressure regulators, mesh filters, and solenoid valves. The overall flow is considered to obey ideal gas laws. The velocity within the porous zone is calculated using Fluent’s physical velocity formulation (Ansys Fluent Theory Guide, 2021). A two-equation SST $k-\omega$ model has been utilized for the entire framework to model the turbulent flow within the domain. Note that the adsorption process is temperature-dependent and strongly exothermic. Hence, the thermal energy is simultaneously conserved here to account for the underlying physics. The fluid phase and adsorbent solid are considered here to be in thermal equilibrium since both the zeolite particle sizes and the flow velocities are small. Energy is also conserved in the wall material of all the columns and pipes through conjugate heat transfer analysis. Correspondingly, the different conservation equations are written as follows.    

\subsubsection{Mass conservation equation}
The mass conservation equation under the common framework \cite{fluent2011ansys} is given as
\begin{equation}
    \frac{\partial (\epsilon_b \rho_g)}{\partial t} + \nabla . (\epsilon_b \rho_g \vec{v}) = \epsilon_b S_m,
    \label{tab:3}
\end{equation}
where $\epsilon_b$ is the porosity (applicable only in porous zones), $\rho_g$ is the fluid density, and $\vec{v}$ is the fluid physical velocity. $S_m$ is the volumetric source term that is applied only in the zeolite columns, and it accounts for the contribution of adsorbed/desorbed gases to the overall mass conservation in the gas phase. This mass source/sink term is given as
\begin{equation}
    S_m = -(1 - \epsilon_b)\rho_p \sum_{i}M_i\frac{\partial q_i}{\partial t}.
    \label{tab: b}
\end{equation}
Here, $\rho_p$ is the adsorbent particle density ($kg/m^3$). $q_i$ is the quantity of the $i^{th}$ gas species adsorbed (mol/kg) and $M_i$ is its corresponding molecular weight. The procedure for calculating the term, ${\partial q_i}/{\partial t}$, is discussed separately in sub-section 3.2.6. 

\subsubsection{Momentum conservation equation}
The momentum conservation equation for the gas-phase \cite{ben2018carbon, fluent2011ansys} in the whole domain is given as 
\begin{equation}
    \frac{ \partial(\epsilon_b \rho_g \vec{v}) }{\partial t} + \nabla.(\epsilon_b\rho_g \vec{v} \vec{v} ) = -\epsilon_b \nabla P + \nabla.(\epsilon_b \mu \nabla.\vec{v}) +  S_f
  \label{tab:12}
\end{equation}
Here, $P$ is the pressure, and $\mu$ is the fluid viscosity. $S_f$ is the momentum drag term that is relevant only to the porous media. In the adsorbent columns, it is specifically calculated using Ergun's equation for packed beds \cite{Ergun} as 
\begin{equation}
    S_f = -\Bigg(\frac{\epsilon_b^2 \mu}{\kappa} \vec{v} + \frac{\epsilon_b^3 C_2}{2} \rho_g |\vec{v}| \vec{v} \Bigg).
    \label{tab:15}
\end{equation}
${1}/{\kappa}$ is the inverse of solid media's permeability and is also referred as the viscous resistance ($ 1/m^2$). C$_2$ is the inertial resistance ($1/m$). For a packed bed made of uniform particles with diameter, $d_p$,  these viscous and inertial resistances are given as 
\begin{equation}
    \frac{1}{\kappa} = \frac{150\,(1-\epsilon_b)^2}{d_p^2\,\epsilon_b^3},
    \label{tab:16}
\end{equation}
\begin{equation}
    C_2 = \frac{3.5\,(1-\epsilon_b)}{d_p\,\epsilon_b^3}.
    \label{tab:17}
\end{equation}

The above $S_f$ terms get modified for other porous zones like pressure regulators, solenoid valves, and mesh filters. In these zones, only the inertial resistance part is considered, and $S_f$ takes the simple form,

\begin{equation}
    S_f = \frac{C_2}{2} \rho_g |\vec{v}| \vec{v}.
    \label{tab:18}
\end{equation}

In the case of the pressure regulator, the $S_f$ term should restrict the exit pressure of the flow to the set pressure value ($P_{set}$) when the reservoir pressure ($P_{res}$) is higher than ($P_{set}$). Hence, the purple region representing the pressure regulator in Fig.~\ref{fig:mesh_dtwin} should enable a total pressure drop of $\Delta P_{PR} = \Delta P_{ln}+\Delta P_{dry}$, wherein $\Delta P_{ln} = P_{res} - P_{set}$ and $\Delta P_{dry}$ is the contribution from the desiccant dryer evaluated as $0.03 \times (P_{res} - \Delta P_{ln})$. From the purview of the porous approximation, this pressure drop of $\Delta P_{PR}$ can be achieved when the $C_2$ value is derived according to the expression given below.

\begin{equation}
    \frac{\Delta P_{PR}}{L_d} = \frac{C_2}{2} \rho_g V^2.
    \label{eq:dpr}
\end{equation}

Here, $L_d$ is the length of the purple `PR' domain in the simulation geometry. Note that the $\Delta P_{ln}$ is set to zero if $P_{res}$ is not greater than $P_{set}$. 

In the case of solenoid valves, first, the ratio between the upstream ($P_1$) and downstream ($P_2$) pressures is used to determine if the flow within the valve is subsonic or supersonic. For subsonic flow conditions, i.e., $P_2 \geq P_1/2$, the manufacturer specified equation \cite{asco} for the pressure drop across the valve is given as

\begin{equation}
 \Delta P = \frac{Q_N^2 \,(SG)_N\,T_1}{K_v^2 \, 514^2\,P_2}
  \label{tab:20}
\end{equation}

Here, $(SG)_N$ is the specific gravity of the gas in relation to air, $T_1 ( \text{in} ^{\circ} C)$ is the fluid temperature at the valve inlet, $Q_N$ is the volumetric flow across the valve, and $K_v$ is the valve flow coefficient. In the current numerical implementation, $Q_N$ is estimated at an interior face next to the valve.

In the case of supersonic flow, the flow within the valve gets choked, and a modified upstream pressure,  $P_{th}$, has to be recalculated using the expression,

\begin{equation}
 P_{th} = \frac{ Q_N \sqrt{(SG)_N . T_1}}{257. K_v}
  \label{tab:21}
\end{equation}

The pressure drop, $\Delta P$, in such cases is calculated as $\Delta P = P_1 - P_{th}$. With the pressure drop thus obtained, the inertial resistance coefficient, $C_2$, is evaluated using Eq.~\eqref{eq:dpr}. This, in turn, helps with the evaluation of the $S_f$ term using Eq.~\eqref{tab:18}.

Finally, the pressure drop across the mesh filter \cite{WU20053008} is calculated as 

\begin{equation}
 \frac{\Delta P_{mesh}}{L} =  \frac{f_k\, S_v} {6}  \frac{(1-\epsilon_m)} {\epsilon_m^3} \rho V^2
  \label{tab:12}
\end{equation}
where $S_v$ (=40000) is the surface area per unit volume of the solid phase  and $\epsilon_m$ (=0.4) is mesh porosity. The friction coefficient, $f_k$, is given as
\begin{equation}
 f_k = \frac{250(1-\epsilon_m)} {Re_m}  +  \left[\frac{1.69(1-\epsilon_m)} {Re_m}\right]^{0.071}
  \label{tab:12}
\end{equation}
Here, the flow Reynolds number, $Re_m$, is defined as
\begin{equation}
 Re_m = (1-\epsilon_m) \frac{6\rho\,V} {S_v \mu}
  \label{tab:12}
\end{equation}

\subsubsection{Energy conservation within the flow domain}
As mentioned earlier, the inclusion of energy conservation in the digital twin of a PSA plant is essential for two reasons: 1) the exothermic nature of the adsorption process and 2) the sensitivity of the adsorption process to the temperature changes in the domain. Hence, the effects of heat generation in the adsorbent bed and its exchange with the surroundings must be carefully estimated. In this regard, one must account for three different entities: the gas phase, the adsorbent particles, and the column walls. Fortunately, the small sizes of the porous particles and the low gas velocities allow for the consideration of local thermal equilibrium between the gas phase and solid particles. Thus, we solve a single equation (shown below) for conserving energy in the flow domain, wherein an average temperature is defined in all the porous regions \cite{fluent2011ansys}.
\begin{equation}
    \frac{\partial [\epsilon_b \rho_g E_g + (1-\epsilon_b)\rho_p E_s]}{\partial t} + \nabla.(\epsilon_b(\rho_g E_g+P) \vec{v}) = \nabla.\Big[\epsilon_b (k_{eff} \nabla T - \sum_{i} h_i\vec{J_i} + \mu(\nabla \vec{v}). \vec{v}) \Big] + \epsilon_b S_E.
    \label{tab:13}
\end{equation}

The above expression includes the effects of enthalpy transport due to species diffusion, multi-component diffusion, and thermal diffusion. Here, the buoyancy effects are neglected. $E_g$ is the total gas energy, and $E_s$ is the total adsorbent particle energy. $h_i$ is the sensible heat energy, and $\vec{J_i}$ is the diffusive heat flux of the $i^{th}$ gas component. The parameter, k$_{eff}$, is the effective thermal conductive of the bed and is expressed as
\begin{equation}
    k_{eff} = \epsilon_b k_g + (1-\epsilon_b)k_s.
    \label{tab:14}
\end{equation}
$k_g$ and $k_s$ are the thermal conductivities of the gas media and the solid particles, respectively. The source term, $S_E$, models the exothermic/endothermic behaviour of the adsorption/desorption process in the zeolite columns, and it is defined as
\begin{equation}
    S_E = (1 - \epsilon_b)\rho_p \sum_{i}\Delta H_i \frac{\partial q_i}{\partial t},
    \label{tab: c}
\end{equation}
where $\Delta H_i$ is the heat of adsorption of a given gas species, $i$.
Note that the expression in Eq.~\eqref{tab:13} simplifies to the pure gas phase energy conservation equation in the clear fluid (non-porous) zones. 

Since the column/pipe walls form the intermediary for heat exchange between the flow media and the ambient atmosphere, a conjugate heat transfer analysis is invoked here wherein Eq.~\eqref{tab:13} is solved in conjunction with the wall energy conservation equation described in the following section.

\subsubsection{Energy conservation at the walls}
In the digital twin model, conserving the transient energy balance in the walls is essential for accurately predicting the adsorption phenomenon. The modes of heat transfer here include 1) convection of heat from the internal flow domain to the wall, 2) heat conduction within the wall, and 3) heat convection between the wall exterior and the ambient atmosphere. Since a conjugate heat transfer analysis is involved here, the first mode does not need any explicit treatment. The second mode, i.e., the transient conduction in the walls, is evaluated by solving the basic conservation equation,
\begin{equation}
    \rho_sc_s\frac{\partial T_s}{\partial t}  = \nabla.\Big[k_s \nabla T_s].
    \label{tab:ecw_cond}
\end{equation}
Here, $\rho_s$, $c_s$ and $k_s$ are the density, specific heat and thermal conductivity of the wall material, respectively. $T_s$ represents the temperature distribution within the solid.

Finally, the convection heat transfer between the wall and the outside ambient (the third mode) is solved using the expression,
\begin{equation}
      \rho_sc_s\frac{\partial T_{co}}{\partial t} =h_{\infty}a_{ow}(T_{co}-T_{\infty}).
    \label{tab: c}
\end{equation}

$T_{co}$ is the temperature at the column's exterior surface. $h_{\infty}$ is the convective heat transfer coefficient between the wall exterior and ambient atmosphere, $a_{ow}$ is the outer surface area of the column, and $T_{\infty}$ is ambient temperature. Here, $h_{\infty}$ has been assumed to have a constant value of 8 $W/m^2K$.

\subsubsection{Species conservation equation}
The last set of conversation equations in the present CFD model corresponds to the prominent gas species of the feed air. In the current formulation, the feed air is considered to be a dry gas (without any water molecules) consisting of only nitrogen, oxygen, and argon in the molar percentage ratio of 78:21:1. Note that argon has also been included in the species list as it undergoes varying levels of adsorption in the zeolite. Correspondingly, a fourth dummy gas species, i.e., helium, has been used to compensate for mass/volume imbalance arising from the arithmetic round-offs during the computations. The scalar transport equation associated with the conservation of the species can be written as \cite{ben2018carbon} 
\begin{equation}
    \frac{\partial (\epsilon_b \rho y_i)}{\partial t} + \nabla .(\epsilon_b\rho y_i v) = \nabla.(\epsilon_b D_{disp,i} \nabla y_i) + \epsilon_b S_{si},
    \label{tab:9}
\end{equation}
where $y_i$ is the mass fraction of the $i^{th}$ species. $D_{disp,i}$ is the mass dispersion coefficient of a particular species, $i$, which is evaluated \cite{ruthven1984principles} as  
\begin{equation}
D_{disp,i}=0.7D_{m,i} + 0.5\,|\vec{v}|\,d_p. \label{eq:disp}
\end{equation}

$D_{m,i}$ is the effective molecular diffusion coefficient of an individual species in multi-component gas mixtures, and it is calculated using the Maxwell-Stefan equations \cite{bird2007} as

\begin{equation}
D_{m,i} = \frac{(1-y_i)}{\sum_{j \neq i}\frac{y_j}{D_{ij}}} 
\end{equation}
The binary diffusion coefficient, $D_{ij}$, is calculated here using Fuller equation \cite{fuller1965comparison} as shown below.

\begin{equation}
D_{ij}=0.01013\frac{T^{1.75}\sqrt{\frac{1}{M_{i}}+\frac{1}{M_{j}}}}{P_{abs}[(\sum_v)_i^{\frac{1}{3}}+(\sum_v)_j^{\frac{1}{3}}]^2}
\end{equation}
Here, $T$ is the gas temperature, $P_{abs}$ is the absolute pressure, and $\sum_v$ is gas diffusion volume. $S_{si}$ is the source/sink term associated with the adsorption/desorption process of the $i^{th}$ species and is given as
\begin{equation}
    S_{si} = -(1 - \epsilon)\rho_p M_i \frac{\partial q_i}{\partial t}
    \label{tab: a}
\end{equation}
The coupled conservation equations mentioned thus far ultimately need the local rate of adsorption values, i.e. $\partial q_i/\partial t$, of various gases to evaluate the overall system dynamics. The procedure for estimating these values is discussed in the following sub-section. 

\subsection{Estimation of the adsorption kinetics}
The adsorption/desorption mass transfer essentially depends on the adsorption capacity of the solid and the characteristics of the adsorbing gas species. While describing these features in a true microscopic sense may involve sophisticated computations, they can often be emulated via simple models from an engineering perspective. Such simple models include the Pseudo first-order model (PFO), the Pseudo second order (PSO) model, the Elovich model and the Intra-particle diffusion (IP) model. In the present study, we use the Linear Driving force (LDF) model to quantify the adsorption kinetics as follows.  
\begin{equation}
    \frac{\partial q_i}{\partial t} = k_{Li}(q^*_i - q_i).
    \label{tab:20}
\end{equation}

Here, $k_{Li}$ is the overall mass transfer coefficient of the species, and $q^*_i$ is the maximum adsorption capacity of the zeolite for a particular species, $i$, at equilibrium. $q_i$ is the actual amount of species, $i$, adsorbed in the zeolite. $q^*_i$ depends on the equilibrium pressure and temperature, and various isotherm models such as Langmuir, Langmuir-Freundlich, Toth, and Brunauer, Emmett, and Teller (BET) can be used to represent the data. Among these, the simplest Single-Site multi-component Langmuir (SSL) isotherms have often been used in the literature to calculate $q^*_i$. This isotherm model assumes
\begin{itemize}
\item a homogeneous distribution of the gas molecules on the surface.
\item that the adsorbent has a fixed number of adsorption sites with identical energies.
\item that each site can accommodate only a single molecule, and there are no interactions between the adsorbed molecules. 
\end{itemize}

Owing to the present consideration of multi-component mixtures, an extended version of the SSL model, as shown below, has been utilized in the present work.
\begin{equation}
    q^*_i = \frac{q_{i}^{sat}b_iP_i}{1 + \sum_{j=1}^{n_{comp}} b_jP_j },
    \label{tab:21}
\end{equation}
$q_{i}^{sat}$ is the saturated solid-phase concentration of the $i^{th}$ component. $b_i$ is the equilibrium constant which follows the Van 't Hoff equation given as
\begin{equation}
    b_i=b_{0,i}\,exp^{\left(-{\Delta H_i}/{RT}\right)},
   \label{tab:22}
\end{equation}
where $\Delta H_i$ is the enthalpy of adsorption. Table~\ref{tab:adsorp_prop} show the values of $q_{i}^{sat}$, $b_{0,i}$, and $\Delta$H for the individual gas species on the zeolite 13X material. These values have been obtained by fitting the empirical data as per the thermodynamically correlated SSL model. More details about the adsorption isotherm can be found in the supplementary information.

\begin{table}[!h]
\renewcommand{\arraystretch}{1.2}
\begin{tabular}{|c|c|c|c|c|} 
  \hline
  {Species} & {$q_{i}^{sat}$} & {$b_{0,i}$} & { $\Delta$H} \\
  \hline
    \ & {mol/kg} & {1/Pa} & {J/mol} \\
  \hline \hline
   \textbf{$Ar$} & 3.252 & 1.58767e-09  & -13161.89 \\
   \hline
    \textbf{$O_2$} & 3.252 & 2.02236e-09 & -12685.50 \\
   \hline
   \textbf{$N_2$} & 3.252 & 5.04224e-10 & -19085.62  \\
   \hline 
 \end{tabular}
\centering
\caption{\label{tab:adsorp_prop} 13X adsorbent properties for individual gas species}
\end{table}

In Eq.~\eqref{tab:20}, the overall mass transfer coefficient, $k_{Li}$, is estimated by accounting for the resistances that oppose species transfer between the gas and the solid phases. For a packed bed of adsorbent particles, three types of mass transfer resistances typically manifest; they are film, macro-pore, and micro-pore resistances. For air separation using 13X zeolite, the molecular diffusion in the macro pores is known to control the mass transfer. The expression for this macro-pore resistance \cite{haghpanah2013multiobjective} is given as 

\begin{equation}
   k_{Li} = \frac{60 \epsilon_p D_{p, i}}{d_{p}^2} \left(\frac{C_i}{q^*_i}\right) 
    \label{tab:23}
\end{equation}
In the above expression, $C_i$ (in mol/m$^3$) is the gas-phase concentration of a particular species and is evaluated from the ideal-gas consideration as
\begin{equation}
   C_{i} = \frac{y_i P} {RT},
    \label{tab:23}
\end{equation}
where $y_i$ is the species molar fraction and $R$ is the universal gas constant. $\epsilon_p$ is the porosity of the adsorbent particle and is specified as a constant value of 0.34. The macro-porous diffusion, $D_{p,i}$, is given \citep{Dantas} as  

\begin{equation}
    {D_{p,i}}= {D_{m,i}}/\tau 
    \label{tab:25}
\end{equation}

Here, $D_{m,i}$ is the molecular diffusivity. $\tau$ represents the tortuosity factor of the binder pore, and it usually varies between 2 and 5 \cite{ruthven1984principles}. In the current work, a constant tortuosity value of 2.5 has been utilized.

\subsection{Material properties}

In the present digital twin, the porous medium within the zeolite tank is represented as a collection of mono-disperse spheres, with radially varying bed porosity, $\epsilon_b$, defined as \cite{MUELLER, deklerk} 

\begin{equation}
  \epsilon_b(r)=\begin{cases}
    2.14 z^2 -2.53 z +1, & {z \leq 0.637}.\\
    \epsilon_{bo} + 0.29 e^{-0.6z} cos[2.3 \pi(z-0.16)]+0.15 e^{-0.9z}, & {z > 0.637}.
  \end{cases}
\end{equation}

Here, the scaled distance from the wall, $z$, is defined as $z = (R-r)/{d_p}$. $\epsilon_{bo}$ is the mean bed porosity (= 0.363), $R$ is the adsorbent column radius, and $r$ is the radial position from the cylinder axis. The above description of radial porosity allows the current model to emulate the wall channelling effects that significantly influence the heat and mass transfer in the columns. The relevant thermo-physical properties of both the column material (SS304) and the adsorbent (13X zeolite), listed in Table~\ref{tab:solid}, have been assumed to be constant, i.e., invariant to the change in process conditions.

\begin{table}[ht!]
\renewcommand{\arraystretch}{1.2}
\begin{tabular}{|c|c|c|}
 \hline \hline
  Property & Stainless steel (SS304) & UOP-HP 13X \\
  \hline
   Density (kg/m$^3$) & 8030 & 1133.43 \\  
  \hline
   Specific heat (J/kg\,K) & 502.48 & 1138   \\
   \hline
   Thermal conductivity (W/m\,K)  & 16.27 &  0.08 \\ 
  \hline \hline 
\end{tabular}
\centering
\caption{\label{tab:fluent axy} Thermo-physical properties of the cylinder material and the adsorbent} \label{tab:solid}
\end{table}

In the case of gas mixtures, various thermo-physical properties, such as specific heat, thermal conductivity, and viscosity, have been evaluated using the local composition of the gas. The mixture-gas density is obtained using the ideal gas law as 
\begin{equation}
    \rho_g = \frac{P_{abs}}{RT\sum_{i} \frac{y_i}{M_i}}
\end{equation}

The specific heat of the gas phase has been obtained using the simple mixing law, whereas the viscosity and the thermal conductivity have been evaluated using mass-weighted mixing law \cite{fluent2011ansys} as  

\begin{equation}
    c_p=\sum_{i}(Y_ic_{p,i}), \,
    \kappa =  \sum_{i} \frac{x_i\kappa_i}{\sum_{j} x_j \phi_{ij}}, \; \& \; \mu = \sum_{i} \frac{x_i\mu_i}{\sum_{j} x_j \phi_{ij}} 
    \label{tab:19}
\end{equation}
where $x_i$ is the molar fraction and the term, $\phi_{ij}$, is given as
\begin{equation}
    \phi_{ij} = \frac{\Bigg[1 + \Big(\frac{\mu_i}{\mu_j}\Big)^{\frac{1}{2}} \Big(\frac{M_j}{M_i}\Big)^{\frac{1}{4}} \Bigg]^2}{\Bigg[8\Big(1 + \frac{M_i}{M_j} \Big) \Bigg]^{\frac{1}{2}}},
    \label{}
\end{equation}
The properties of the individual species used in the above expression have been listed in Table~\ref{tab:spec_prop}. Here, specific properties of O$_2$ and N$_2$ have been obtained as a polynomial function of temperature as $A + B\, T + C\, T^2 + D\, T^3 + E\, T^4$. The corresponding coefficients are listed in Table~\ref{tab:poly_coeff}.

\begin{table}[!h]
 \begin{adjustbox}{max width=\textwidth}
\begin{tabular}{|c|c|c|c|c|c|} 
  \hline
  {Gas} & {Molecular weight} & {Specific Heat} & {Thermal Conductivity} & {Viscosity}  & {$\sum_v$ {\cite{fuller1965comparison}} }\\
  \hline
    \ & {kg/kmol} & {J/kg-K} & {W/m-K} & {kg/m-s} &   \\
  \hline \hline
   Ar & 39.948 & 520.64 & 0.0158 & 2.125e-05 & 16.2\\
   \hline
   O$_2$ & 31.9988 & polynomial & polynomial & polynomial & 16.3\\
   \hline
   N$_2$ & 28.0134 & polynomial & polynomial & polynomial & 18.5\\
   \hline
   He & 4.0026 & 5193 & 0.152 & 1.99e-05 & 2.67 \\
   \hline 
 \end{tabular}
 \end{adjustbox}
\centering
\caption{\label{tab:spec_prop} Properties of individual gas species}
\end{table}

\begin{table}[ht!]
 \renewcommand{\arraystretch}{1.3}
 \begin{tabular}{|c|c|c|c|c|c|}
			
 \hline  \hline
 \textbf{} & \textbf{A} & \textbf{B} & \textbf{C} & \textbf{D} & \textbf{E} \\
 \hline 
 \textbf{$C_{P_{O_2}}$} & 811.1803 & 0.4108345 & -1.750725e-4 & 3.757596e-8 & -2.973548e-12\\
 \hline
 \textbf{$C_{P_{N_2}}$} & 938.8992 & 0.3017911 & -8.109228e-5 & 8.263892e-9 & -1.537235e-13\\
 \hline
 \textbf{$K_{O_2}$} & 3.921754e-3 & 8.081213e-5 & -1.354094e-8 & 2.220444e-12 & -1.416139e-16\\
 \hline
 \textbf{$K_{N_2}$} & 4.737109e-3 & 7.271938e-5 & -1.122018e-8 & 1.454901e-12 & -7.871726e-17\\
 \hline
 \textbf{$\mu_{O_2}$} & 7.879426e-6 & 4.924946e-8 & -9.851545e-12 & 1.527411e-15 & -9.425674e-20 \\
 \hline
 \textbf{$\mu_{N_2}$} & 7.473306e-6 & 4.083689e-8 & -8.244628e-12 & 1.305629e-15 & -8.177936e-20\\
 \hline 
 \end{tabular}
 \centering
 \caption{Polynomial coefficients for different properties}\label{tab:poly_coeff}
\end{table}

\subsection {Numerical discretization}
Towards completing the digital twin model discussed thus far, the relevant details of the numerical discretization procedures employed here are briefly discussed in this subsection. For the present transient 2D-axisymmetric flow simulations, an implicit pressure-based SIMPLE algorithm was utilized to establish a proper association between the velocity and pressure fields. This approach ensures the overall mass conservation in the domain and helps evaluate the pressure field. Here, the face values of pressure were computed from cell values using a second-order interpolation scheme. The advective terms of all the momentum, energy, and species transport equations have been spatially discretized using a second-order upwind algorithm. At the same time, a least squares cell-based algorithm was utilized for the diffusive terms. In the case of the two-equation SST $k-\omega$ turbulence model, the default first-order upwind algorithm was used for estimating the turbulent kinetic energy and the turbulent dissipation rate. Temporal discretization of all the conservation equations was achieved through a first-order implicit formulation. The adsorption/desorption source terms were modelled using an implicit second-order scheme. All the source terms pertaining to the various conservation equations were incorporated in Ansys-Fluent using User-Defined Functions (UDFs). Also, the equation for the adsorption kinetics was solved at each iteration using the UDFs.    

\subsection {A single column breakthrough study}

\begin{figure}[!ht]
    \includegraphics[width = 14cm]{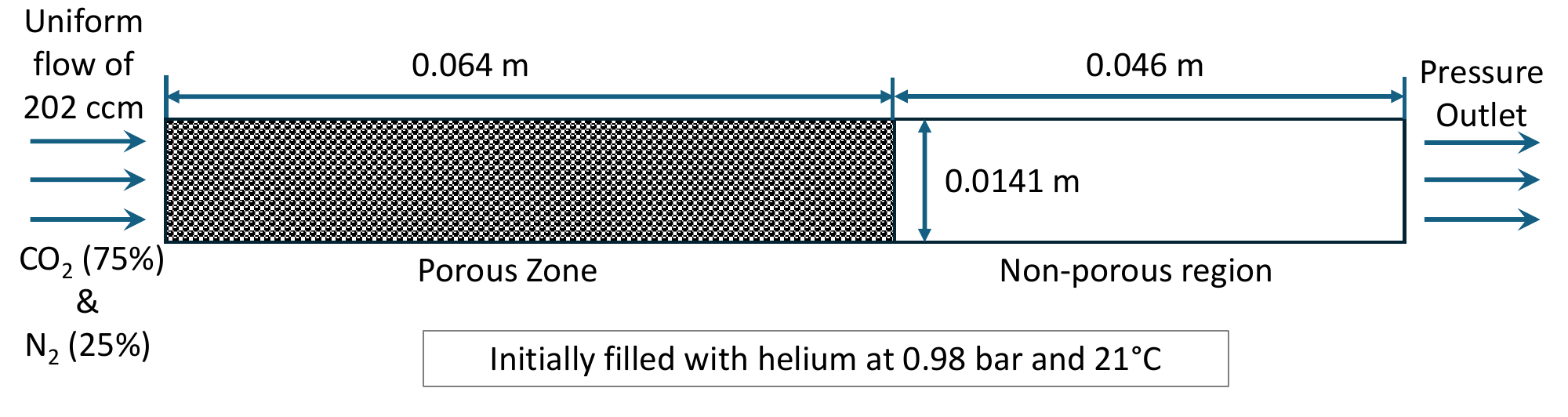}
    \centering
    \caption{Schematic representation of the CO$_2$ - N$_2$ adsorption breakthrough problem}    \label{fig:verify}
\end{figure}

Before assessing the efficacy of the current digital twin, we perform a simple consistency check to gauge the model's ability to predict the fundamental adsorption process. In this regard, we resort to the gas breakthrough experiments performed by Wilkins and Rajendran \cite{Wilkins2019MeasurementOC}. Though their experiments involved both the single gas and binary gas systems, only the binary gas system involving the CO$_2$ - N$_2$ combination has been reproduced here for brevity. The experimental configuration of Wilkins and Rajendran \cite{Wilkins2019MeasurementOC} consisted of a cylindrical column with a diameter of 0.0282 m and a length of 0.064 m. The column was filled with 13X zeolite particles of approximately 1 mm diameter. Initially, the adsorbent bed was saturated with helium gas at 0.98 bar and 21°C. A gas mixture of 75\% CO$_2$ and 25\% N$_2$ was fed into the column at a constant flow rate of 202 ccm, and the composition of the outflowing gas was continuously measured. 

In the current numerical consistency check, we reproduce the configuration of Ramos et al. \cite{FABIANRAMOS}, who have performed 2D axisymmetric simulations of the above binary gas breakthrough process. A schematic representation of the numerical test configuration is shown in Fig. 5. Here, the exit domain is extended to include a non-porous region to mitigate the effects of outflow boundary condition on the breakthrough calculation. Although this test case does not involve boundary manipulations as in the present digital twin, it still helps to verify the implementation of all the conservation equations and their source terms that have been specified via UDFs. Figure~\ref{fig:verify} presents three plots corresponding to CO$_2$ breakthrough, N$_2$ breakthrough, and center-line temperature variation. The numerical results closely match those of Ramos et al. \cite{FABIANRAMOS}, and the deviation from the experimental data of Wilkins and Rajendran \cite{Wilkins2019MeasurementOC} is similar to the numerical data of Ramos et al. \cite{FABIANRAMOS}. With the confidence thus gained from the accurate reproduction of numerical breakthrough data, we now proceed to the full PSA plant simulations and their comparison with the in-house PSA plant's experimental data. 

\begin{figure}[!ht]
    \includegraphics[width = 13.4cm]{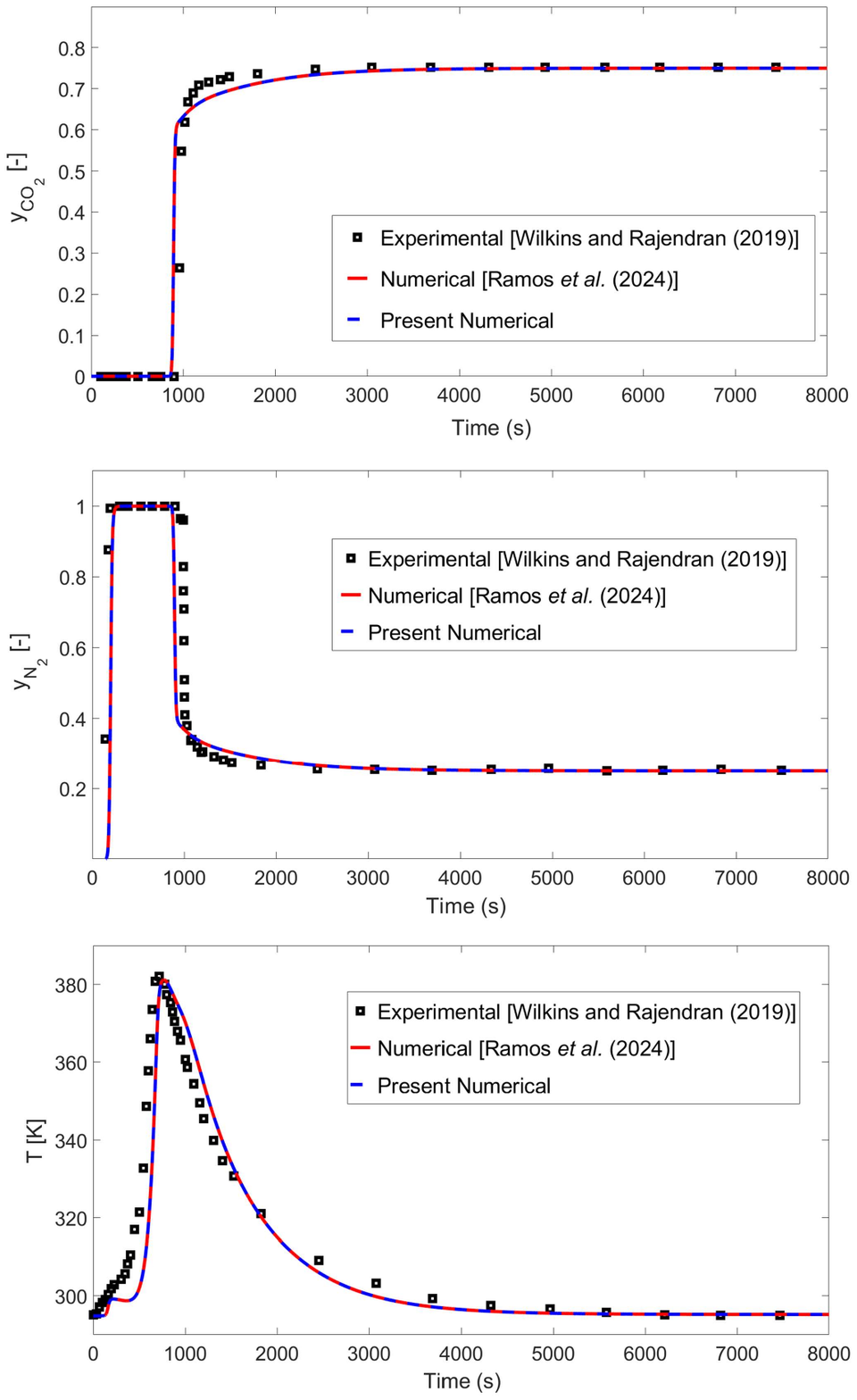}
    \centering
    \caption{A single column test case pertaining to CO$_2$ - N$_2$ breakthrough in 13X zeolite. Top: CO$_2$ breakthrough curve; Middle: N$_2$ breakthrough curve; Bottom: Center-line temperature at the outlet.}    \label{fig:verify}
\end{figure}  

\section{Modelling of oxygen separation using the digital twin model}

We now illustrate the facets of the present digital twin model through a specific case of oxygen separation from air using 13X zeolite. Here, the special focus is on understanding how the device's performance is influenced by the duration of different steps in the cycle, i.e., the pressurization, purge, and equalization times. The current digital twin simulations start from an unpressurized state of the system, which is entirely filled with atmospheric air. Like in experiments, the reservoir is initialized at 9.5 bar of absolute pressure, and its inlet condition (`IN' in Fig.~\ref{fig:mesh_dtwin}) is specified as mass-inlet. The inflow rate was maintained at 0.006341 kg/s, corresponding to the corrected FAD (320 LPM) of the compressor, which includes 12\% loss due to the elevated compressed air temperature (around 140°C) and 15\% purge loss in the dryer (manufacturer-specified). To mimic the real-world operations of the compressor, the inlet flow is turned off (made zero) when the reservoir pressure equals or exceeds 9.5 bar and is re-activated to the above-mentioned value when the pressure falls below 7.5 bar. 

As in the experimental trials, a constant mass flow rate of 4.89 $\times 10^{-4}$ kg/s (22.43 SLPM) was issued from the outlet (`OUT') of the O$_2$ storage tank. The pressure inside the two adsorbent columns and the outlet gas composition were continuously monitored in the simulations and were compared with the experimental data from the in-house prototype plant. The overall sequencing of the steps, according to the modified Skarstorm cycle, was effectuated by modifying the boundary conditions as mentioned in the previous section. Note that during each step change, one end of both columns is always subjected to an abrupt pressure change. For example, when pressurization-1 or pressurization-2 begins, the pressurizing column is exposed to the high-pressure feed air at the bottom, while the depressurizing column suddenly opens up to atmospheric conditions. These scenarios often lead to momentary high Mach number flow within the columns that may restrict the computational time step that can be used for integration. Accordingly, during each such change, the time-step size was systematically increased from $10^{-5}$ s to $10^{-2}$ s to obtain a stable solution. At the beginning of each time step/iteration, appropriate source terms and local adsorption amount were evaluated using UDFs. The numerical convergence of the calculated solution was ensured by demanding a reduction in the scaled residual values to $10^{-4}$ for all equations at each time step. In each simulation trial, the modified Skarstorm cycle was repeated until the output parameters exhibited a Cyclic Steady-State (CSS) behaviour. In most cases, this occurred at around 20 to 25 cycles of repetition.    

\subsection {Grid independence}
\begin{figure}[]
    \includegraphics[width = 14cm]{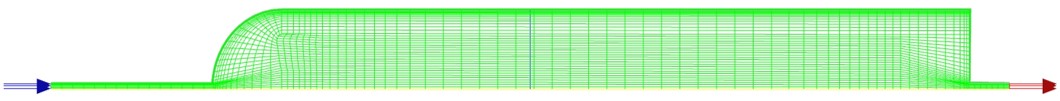}
    \centering
    \caption{A sample grid of the adsorbent column}    \label{fig:grid-ind}
\end{figure}

\begin{figure}[]
    \includegraphics[width = 10cm]{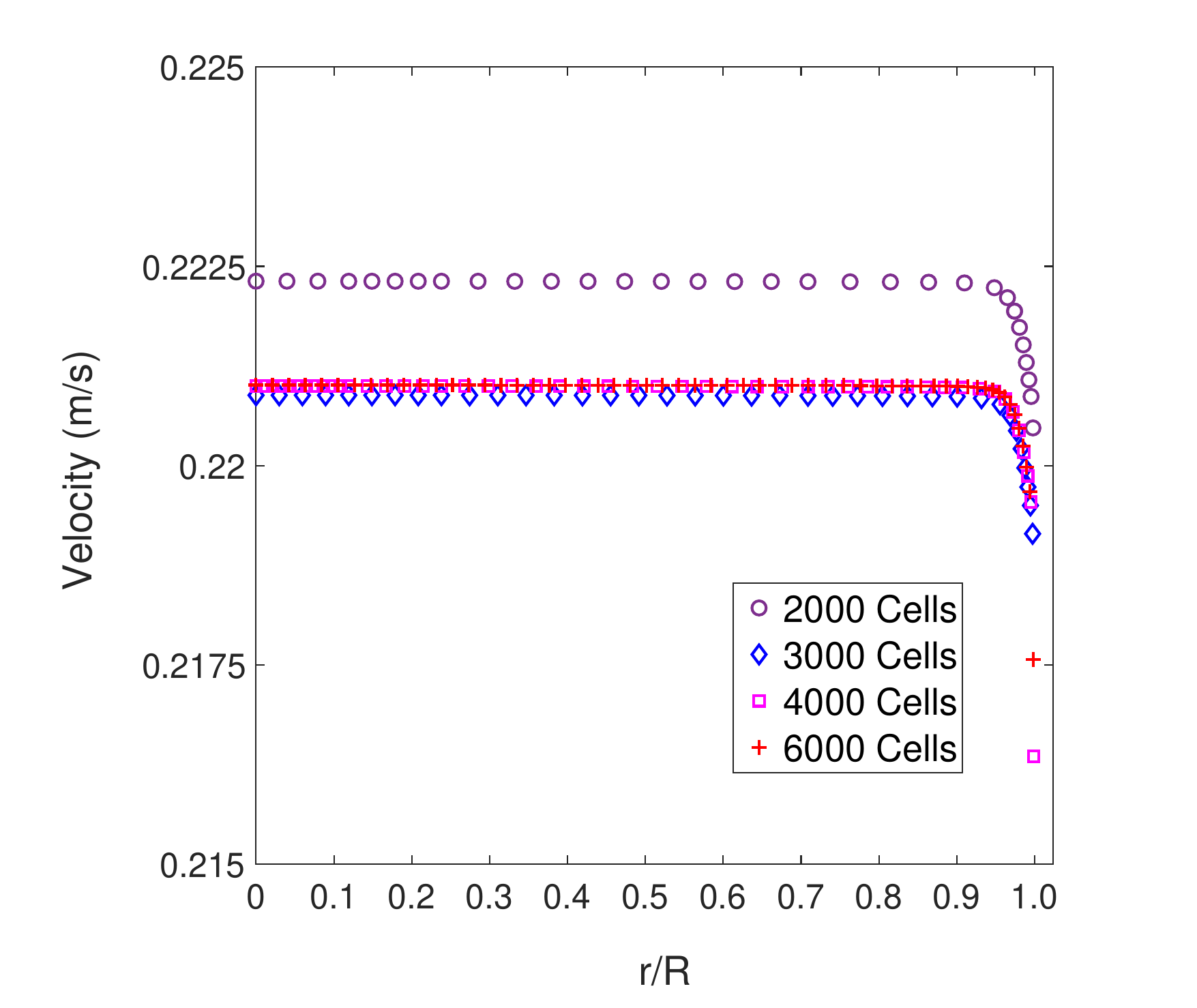}
    \centering
    \caption{Grid independency study.}  \label{fig:vel-ind}  
\end{figure}

Before proceeding to the computations, the optimal grid required for the current digital twin to obtain  accurate solutions at a minimal computational cost is estimated. Since adsorption is the primary focus in any PSA system, the present grid independence study has been particularly focused on the adsorbent columns. The primary objective here is to optimize the grid for the most challenging high-speed flow scenario within the column, which is the start of the pressurization step that involves high-pressure air suddenly entering the (post-equalized) column. In this regard, a stand-alone adsorbent column is considered here with an inlet pressure of 7 bar and an outlet pressure of 4 bar. As shown in Fig.~\ref{fig:grid-ind}, a multi-block structured grid, generated using the ICEM-CFD package, discretizes the adsorbent column of dimensions 930 mm in length and 168.3 mm in diameter. The inlet and outlet pipes are 200 mm and 50 mm long. A C-grid has been utilized along with the refinement of the near-wall region to account for the wall-channelling effects. Transient flow simulations involving four grid candidates, viz. 2000, 3000, 4000, and 6000 quadrilateral cells have been performed. The properties of the adsorbent column and the inlet gas mixture are as specified in the previous section. All relevant parameters, such as flow velocity, temperature, and gas composition, have been monitored at the mid-cross-section (x/L = 0.5) of the zeolite column. However, for brevity, only the velocity profiles have been plotted in Fig.~\ref{fig:vel-ind} since the corresponding inference also holds for other parameters. Figure~\ref{fig:vel-ind} shows the velocity profiles obtained for different grids at the time, $t = 2$ s, from the start of the simulation. Between the grids involving 2000 and 3000 cells, the change in the centreline velocity is around 0.59\%. This change reduces to 0.14\%  when the resolution is increased to 4000 cells. A further increase in cells to 6000 results in a meagre change of 0.09\%. Thus, considering both computational efficiency and accuracy, the grid with 4000 quadrilateral cells has been deployed for the axisymmetric representation of both the adsorbent columns. The other parts of the computational domain have been discretized with similar multi-block structured grids.

\begin{figure}[!ht]
   \centering
    \includegraphics[width=14cm]{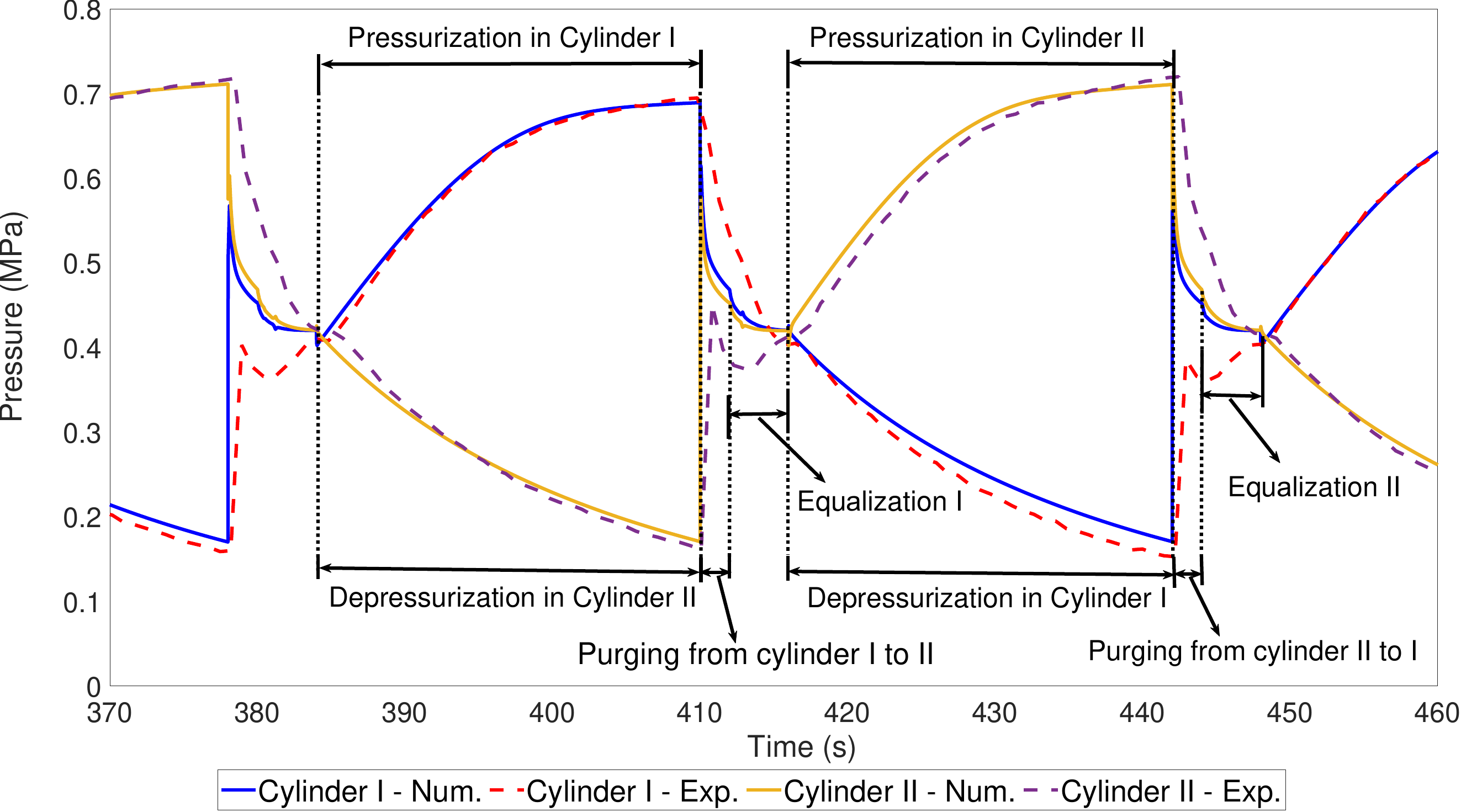}  
    \caption{Transient comparison of adsorbent column pressures for a cycle with t$_{pr}$ = 26s, t$_{pu}$ = 2s, t$_{eq}$ = 4s }
   \label{fig:pres_comp}
\end{figure}


\begin{figure}[!ht]
    \centering
    \begin{tikzpicture}
        \node at (10, 0) {\includegraphics[width=13.25cm]{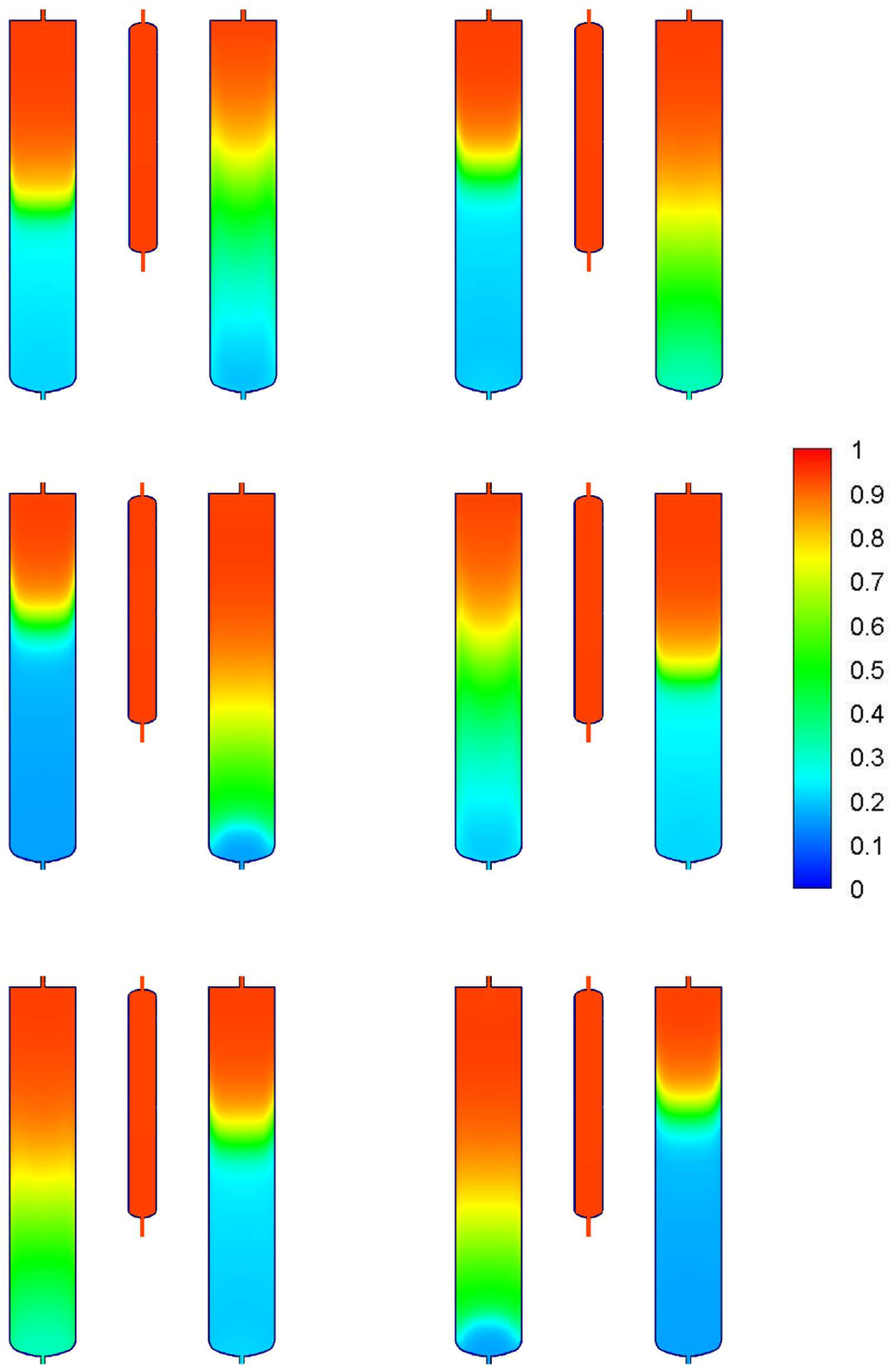}}; 
        \node[anchor=north west] at (6.2, 4) {\textbf{(a)}}; 
        \node[anchor=north west] at (11.8, 4) {\textbf{(b)}}; 
        \node[anchor=north west] at (6.2, -2.1) {\textbf{(c)}}; 
        \node[anchor=north west] at (11.8, -2.1) {\textbf{(d)}}; 
        \node[anchor=north west] at (6.2, -8.3) {\textbf{(e)}}; 
        \node[anchor=north west] at (11.8, -8.3) {\textbf{(f)}}; 
    \end{tikzpicture}
    \caption{Contours of oxygen volume fraction: end of a) Pressurization-1, b) Purge-1, c) Equalization-1, d) Pressurization-2, e) Purge-2, and f) Equalization-2 steps}
    \label{fig:o2_cont}
\end{figure}

\begin{figure}[!ht]
    \centering
    \begin{tikzpicture}
        \node at (10, 0) {\includegraphics[width=13.25cm]{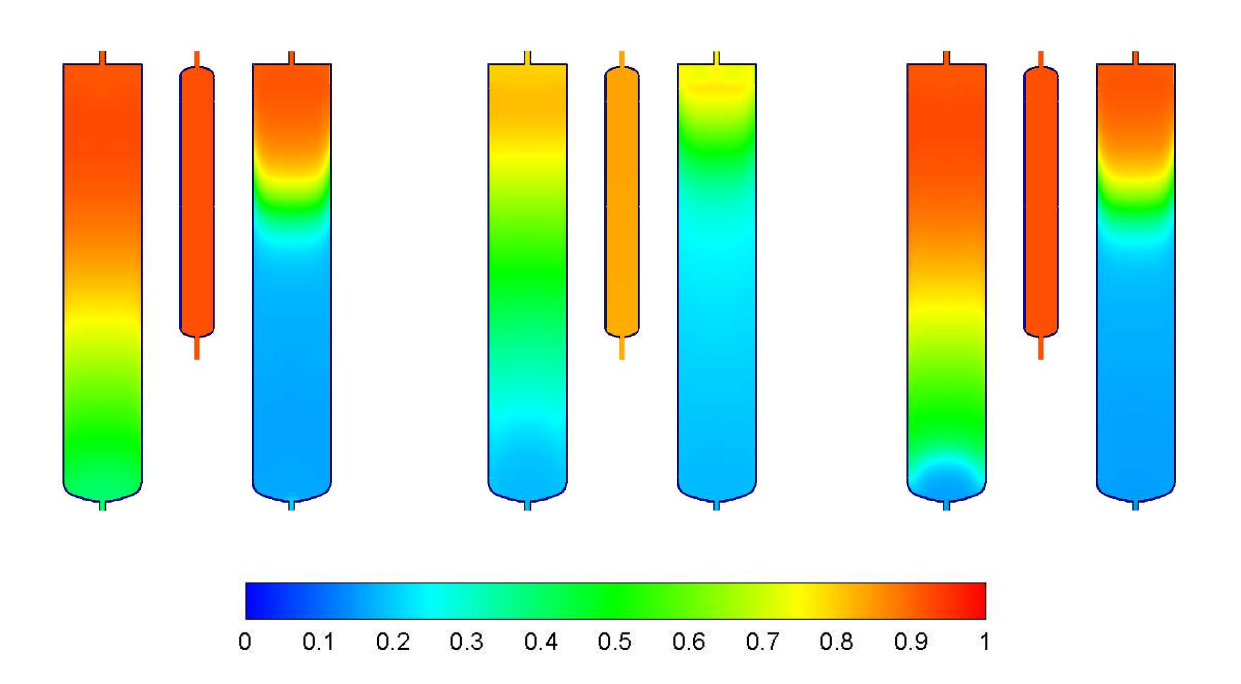}}; 
        \node[anchor=north west] at (5, -1.8) {\textbf{(a)}}; 
        \node[anchor=north west] at (9.6, -1.8) {\textbf{(b)}}; 
        \node[anchor=north west] at (14.1, -1.8) {\textbf{(c)}}; 
    \end{tikzpicture}
    \caption{Contours of oxygen volume fraction at the end of equalization step for a) Top-top, b) Bottom-bottom, and c) Top + Bottom configurations}
    \label{fig:eq_var}
\end{figure}
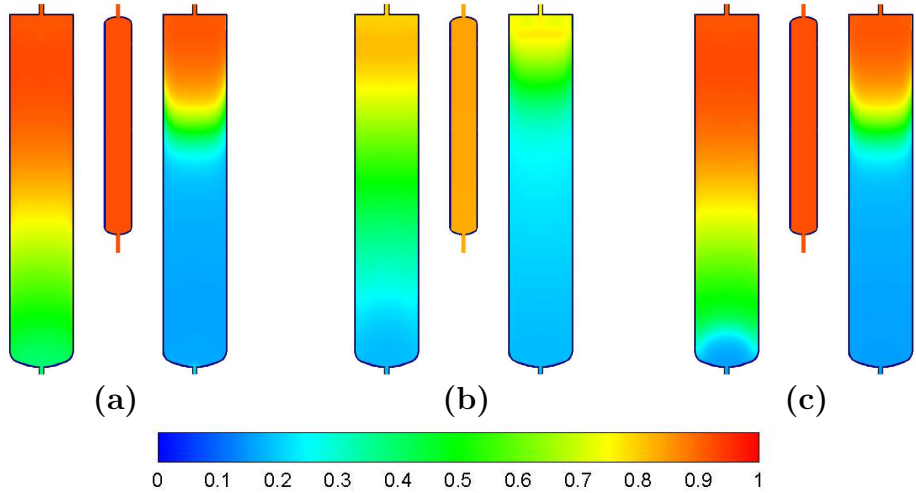

\begin{figure}[!ht]
    \centering
    \begin{tikzpicture}
        \node at (10, 0) {\includegraphics[width=13.25cm]{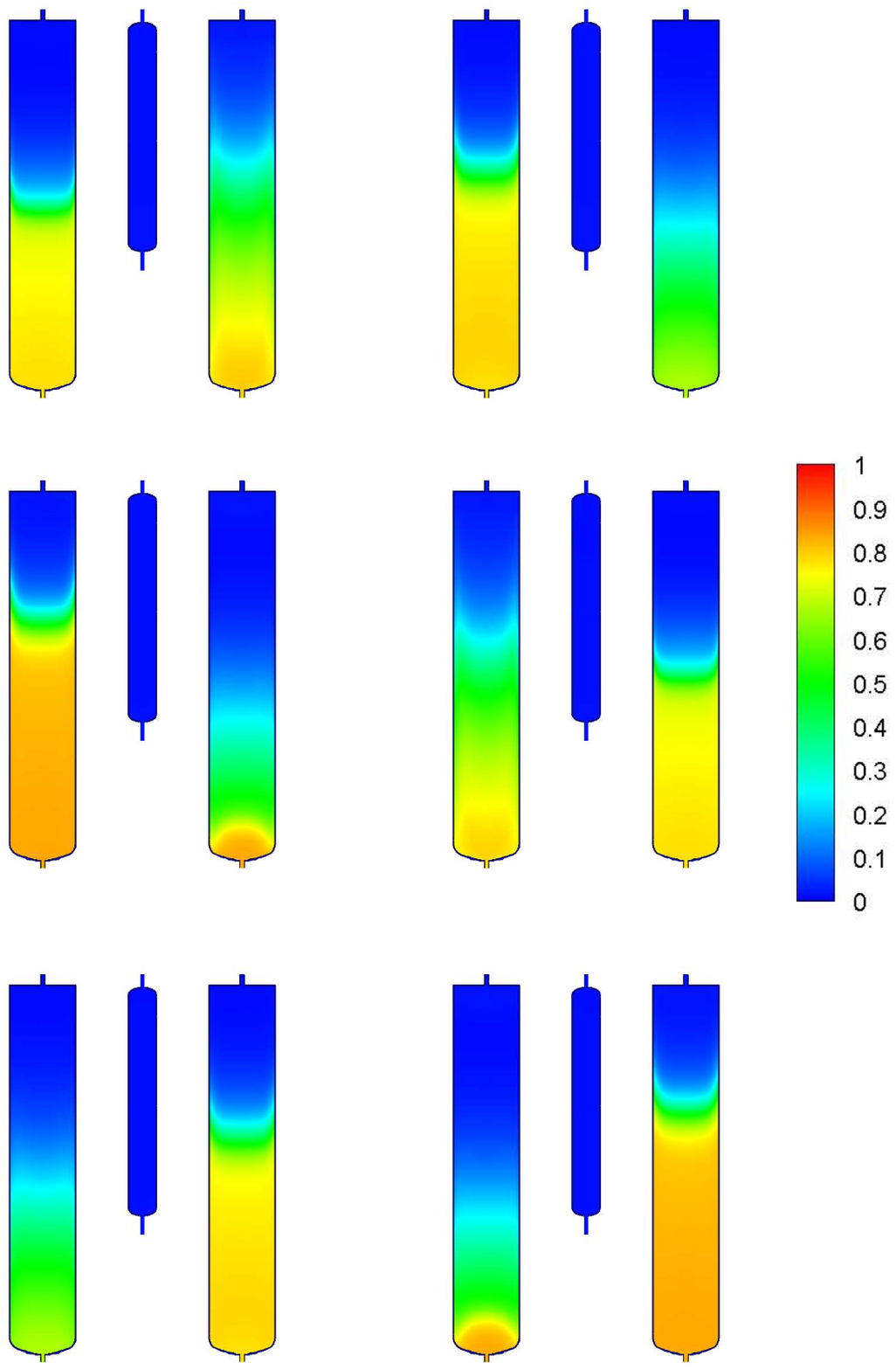}}; 
        \node[anchor=north west] at (6.2, 4) {\textbf{(a)}}; 
        \node[anchor=north west] at (11.8, 4) {\textbf{(b)}}; 
        \node[anchor=north west] at (6.2, -2.1) {\textbf{(c)}}; 
        \node[anchor=north west] at (11.8, -2.1) {\textbf{(d)}}; 
        \node[anchor=north west] at (6.2, -8.3) {\textbf{(e)}}; 
        \node[anchor=north west] at (11.8, -8.3) {\textbf{(f)}}; 
    \end{tikzpicture}
    \caption{Contours of nitrogen volume fraction: end of a) Pressurization-1, b) Purge-1, c) Equalization-1, d) Pressurization-2, e) Purge-2, and f) Equalization-2 steps}
    \label{fig:n2_cont}
\end{figure}

\begin{figure}[!ht]
    \centering
    \begin{tikzpicture}
        \node at (10, 0) {\includegraphics[width=13.25cm]{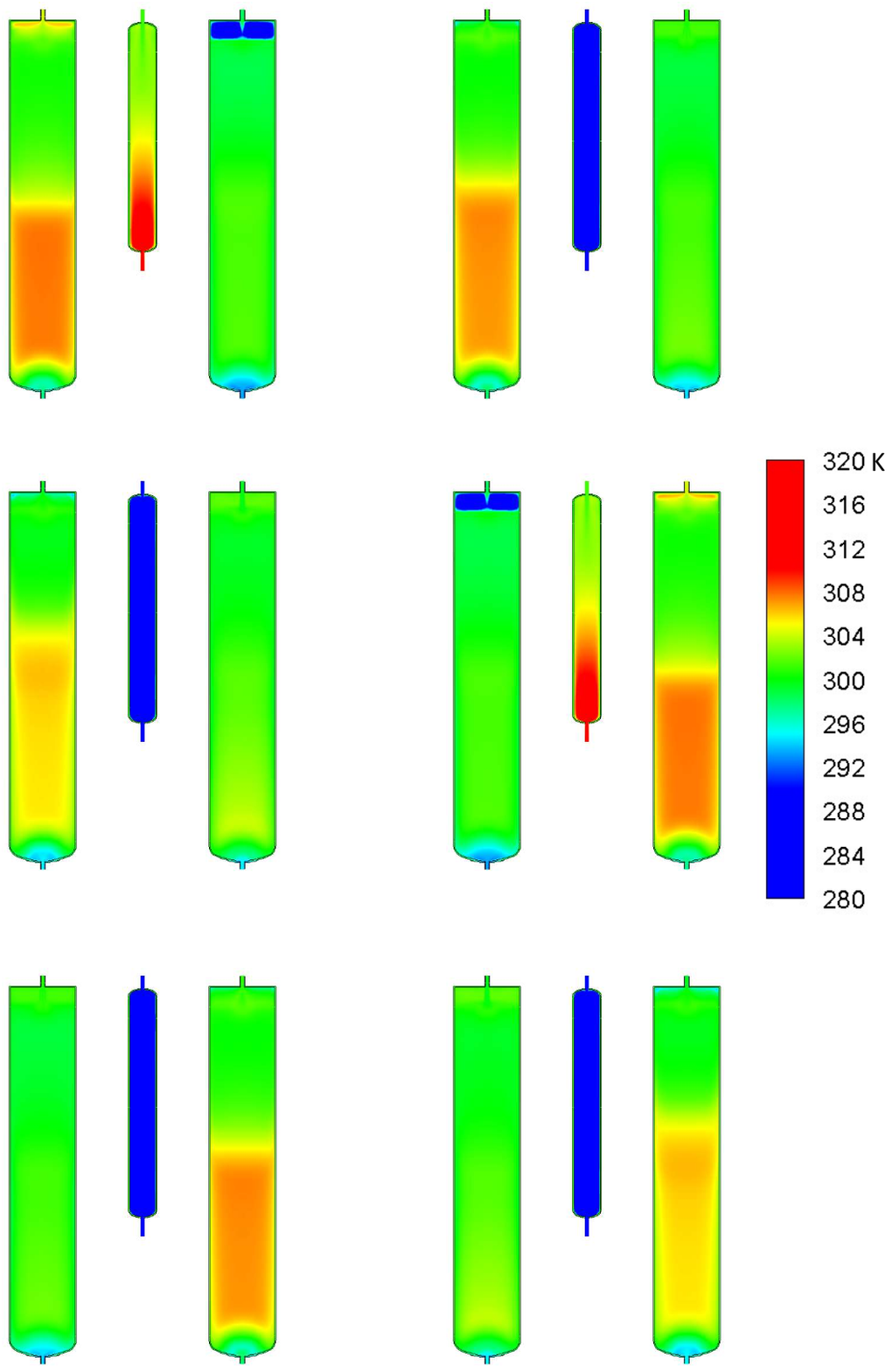}}; 
        \node[anchor=north west] at (6.2, 4) {\textbf{(a)}}; 
        \node[anchor=north west] at (11.8, 4) {\textbf{(b)}}; 
        \node[anchor=north west] at (6.2, -2.1) {\textbf{(c)}}; 
        \node[anchor=north west] at (11.8, -2.1) {\textbf{(d)}}; 
        \node[anchor=north west] at (6.2, -8.3) {\textbf{(e)}}; 
        \node[anchor=north west] at (11.8, -8.3) {\textbf{(f)}}; 
    \end{tikzpicture}
    \caption{Contours of temperature: end of a) Pressurization-1, b) Purge-1, c) Equalization-1, d) Pressurization-2, e) Purge-2, and f) Equalization-2 steps}
    \label{fig:temp_cont}
\end{figure}

\subsection{Features of the modified Skartsorm cycle}

The detailed flow and adsorption behaviour within the PSA system during a single Skarstorm cycle is now described using the present comprehensive digital twin model. For this purpose, the numerical data, particularly the pressure values at the top of adsorbent tanks, have been compared in Fig.~\ref{fig:pres_comp} with the experimental results obtained from the in-house PSA plant. The contours of O$_2$ and N$_2$ volume fractions and temperature fields at the end of different steps are shown in Figs. \ref{fig:o2_cont}, \ref{fig:n2_cont}, and \ref{fig:temp_cont}, respectively. The supplementary video appended to this work provides a complete picture of these fields' variation in a cycle. The cycle used for the present illustration involves a pressurization time (t$_{pr}$) of 26s, purge time (t$_{pu}$) of 2s, and equalization time (t$_{eq}$) of 4s. It is worth remembering that the output flow rate of the oxygen-enriched gas is maintained constant at 4.89 $\times 10^{-4}$ kg/s for all the trials that have been performed. From Fig.~\ref{fig:pres_comp}, a close match between the numerical and the experiment pressure data can be observed, primarily during the pressurization/depressurization process. This close match indicates the high efficacy of the present adsorption/desorption kinetics model in mimicking that actual process. However, there is a notable difference between the experimental and numerical data during the purge and equalization steps. During the purge step, both columns experience sudden changes in the pressure; this is well captured in the numerical data. The experimental curves, owing to the latency in the pressure probes, have a reduced slope and do not reach the same peak as in the numerical simulations. Similarly, during the equalization step, the numerical pressure values in both columns are close to each other. The experimental pressure values, on the other hand, take the whole t$_{eq}$ to equilibrate. Fortunately, they recover well enough to properly predict the pressure changes during the pressurization/depressurization processes again. Thus, there is a repeated consistency between the measured experimental data and the numerical predictions.  

Different contour plots shown in Figs. \ref{fig:o2_cont}, \ref{fig:n2_cont}, and \ref{fig:temp_cont} reveal the various intricate details of the overall PSA process. For the sake of conciseness, only the contours at the end of each step have been represented here. Figure \ref{fig:o2_cont} clearly shows the movement of the oxygen front in both columns during the cycle. Owing to the wall-channelling effect, these fronts are not planar, and they inform the extent to which any step can be extended, i.e., the maximum allowable value for t$_{pr}$, before the purity of the output gas is affected. During the pressurization steps, the lower part of the pressurizing column shows the extent of adsorption saturation in the zeolites as the local O$_2$ volume fraction is closer to the value in atmospheric air. In the desorbing column, however, the leading O$_2$ front moves downward, and there is a significant desorption of N$_2$ in the lower part, which makes the local O$_2$ volume fraction very small. The purge step enables the faster downward movement and dilution of the leading O$_2$ front in the desorbing column. Consequently, a large part of the desorbed nitrogen is blown out of the column. However, during the equalization step, nitrogen reenters this column from the pressurized column since the bottom-bottom equalization is also allowed in the present modified Skarstrom cycle. One can also realize other equalization configurations wherein only the top-top and bottom-bottom equalization paths are utilized. The current digital twin allows for an easy assessment of such alterations by simply modifying the boundary conditions. Figure~\ref{fig:eq_var} shows the O$_2$ concentration at the end of the equalization step for all the three configurations, viz. a) top-top alone, b) bottom-bottom alone, and c) combined top + bottom. The figure clearly shows the dilution of oxygen front in the bottom-bottom case, which has detrimental effects on output purity. The CSS O$_2$ concentration at the outlet is lowest (85\%) for this configuration as compared to top-top (92.93\%) and top + bottom (92.78\%) configurations. Both the latter cases exhibit similar characteristics in terms of the oxygen front, except for a minor re-entry of nitrogen back in the pressurized column for the top + bottom case. The N$_2$ contours in Fig. \ref{fig:n2_cont}, though having complementary contour colours, corroborate all the processes described above. 

The temperature contours in Fig. \ref{fig:temp_cont} reveal interesting information, particularly on the buffer tank's temperature oscillations. During the pressurization and depressurization steps, the clear fluid's temperature at the top of the pressurizing column accordingly increases while that of the depressurizing column decreases. The fluid transferring from the pressuring column to the O$_2$ storage tank increases the latter's temperature. However, during the purge step, there is a sudden decrease in the buffer tank temperature due to the rapid decrease in the tank's pressure. The configuration of the solenoid valves in the present PSA plant allows for backflow from the buffer tank to the desorbing column when the purge valve (SV1) is opened. This results in an abrupt pressure change in both the zeolite columns and the buffer tank. During the equalization process, the low temperature in the buffer tank is retained as it is cut off from the adsorbent columns. With the restart of the pressurization process, the temperature of the buffer tank once again starts to rise. Note that the transient change observed in all the fields is a strong function of the process parameters such as t$_{pr}$, t$_{pu}$, and t$_{eq}$. We now analyze their influence on the overall performance of the device via the output purity.    

\subsection {Effect of pressurization time}
\begin{figure}[!ht]
    \centering
     \includegraphics[width=8.65cm]{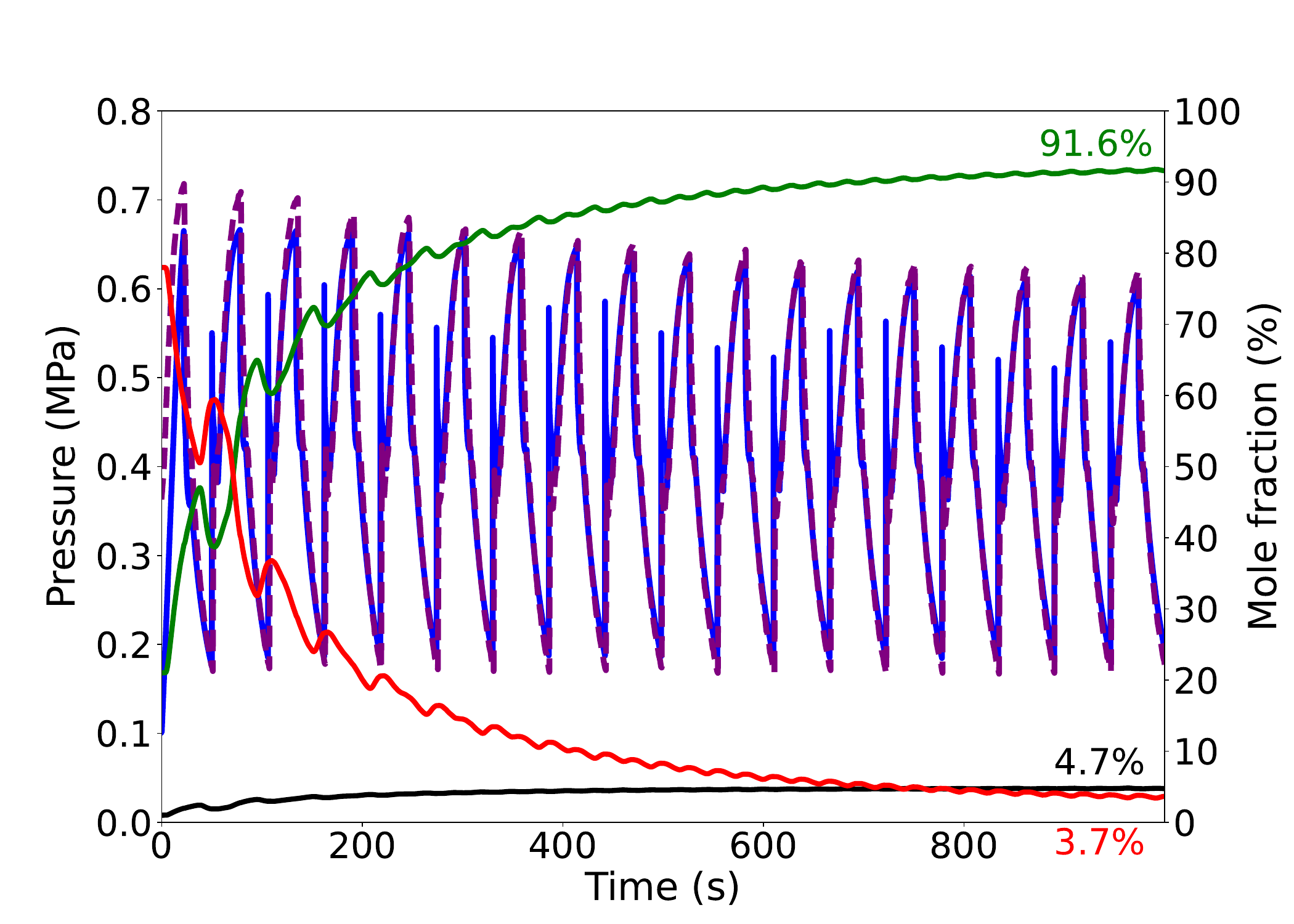}   
     \includegraphics[width=8.65cm]{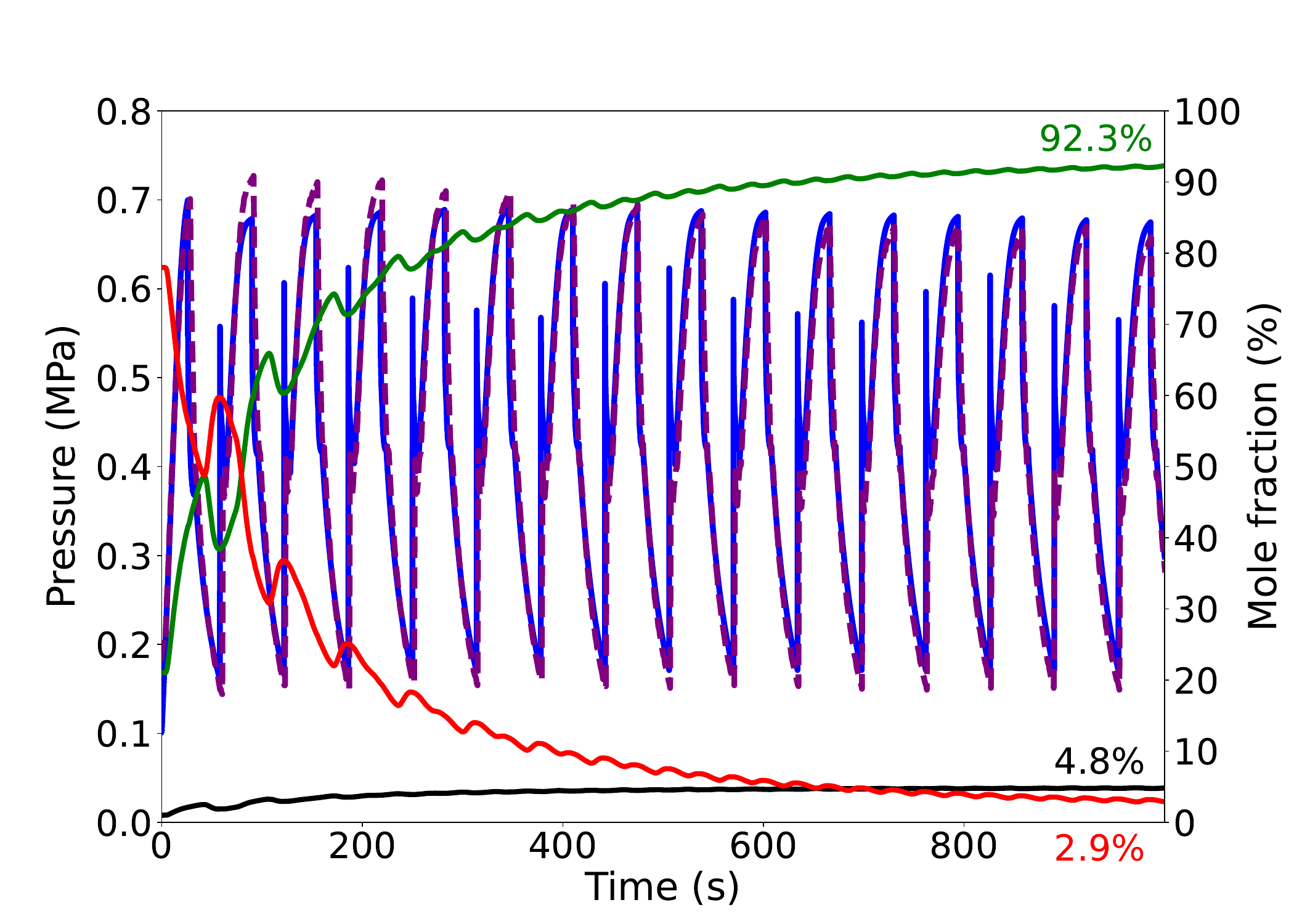} 
     \includegraphics[width=8.65cm]{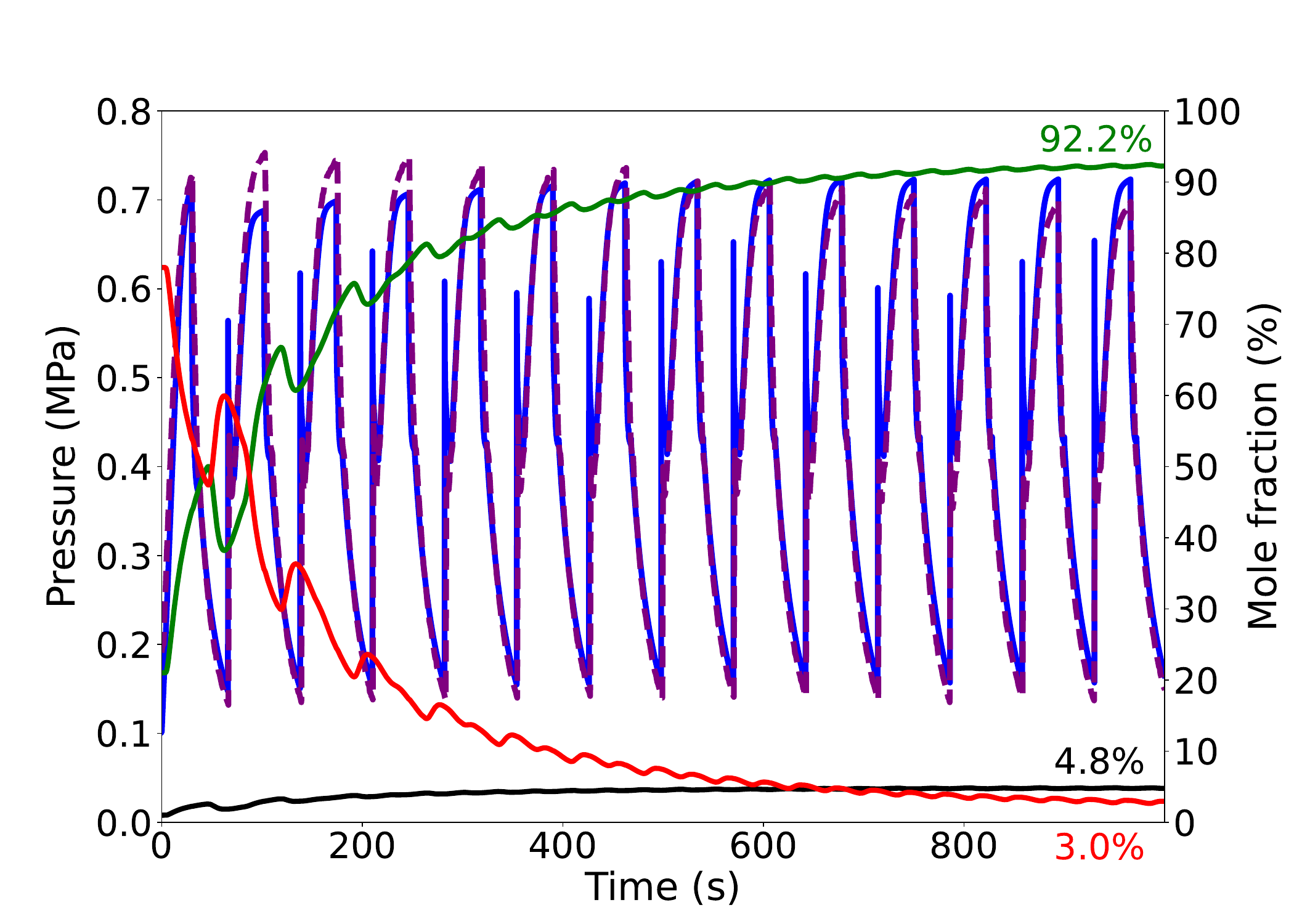} 
     \includegraphics[width=10cm]{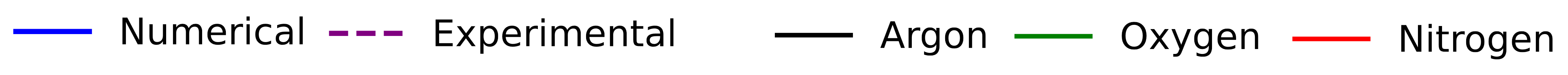}
    \caption{Pressure variation in the adsorbent column and species concentration at the O$_2$ cylinder's exit for t$_{pu}$ = 2s, t$_{eq}$ = 4s, and t$_{pr}$ = (a)22s (b) 26s and (c)30s.}
    \label{fig:Press_cycle}
\end{figure}

\begin{figure}[!ht]
    \centering
     \includegraphics[width=12cm]{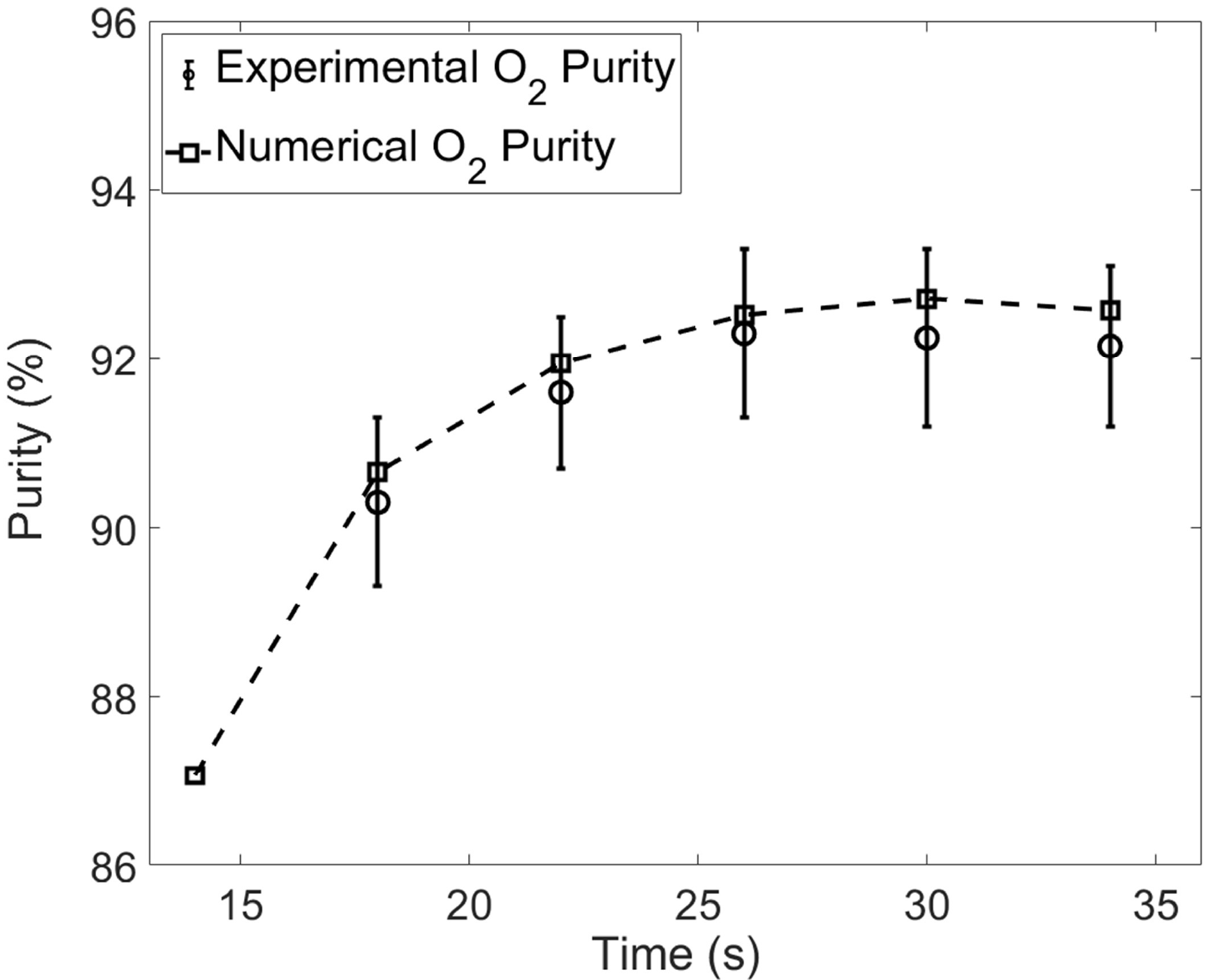}     
    \caption{Effect of pressurization time on purity}
    \label{fig:Pres_curve}
\end{figure}

In a typical PSA plant, optimizing t$_{pr}$ is crucial for both from the perspective of output purity and energy cost. To investigate this, we now vary t$_{pr}$ from 14 s to 34 s, while maintaining t$_{pu}$ and t$_{eq}$ at 2 s and 4 s, respectively. Accordingly, Fig.~\ref{fig:Press_cycle} shows the temporal variation in the adsorbent column's pressure and the species volume fraction at the O$_2$ cylinder's exit for three different t$_{pr}$ viz. 22 s, 26 s, and 30 s. A close match between the numerical and experimental pressure values is observed, except for the spike region during the purging process. The composition of different gases at the device exit shows saturation behaviour at around 1000 s of operation in each case. The influence of t$_{pr}$ is primarily evident in the transient pattern of the peak pressure values in the adsorbent columns. Note that a shorter cycle time, owing to shorter t$_{pr}$, would demand larger flow rates from the air receiver tank on account of frequent pressurization and depressurization of the adsorbent columns. Since the device is powered by a compressor with a constant flow rate, a gradual reduction in the peak pressures of both the air receiver tank and the adsorbent columns can be observed. Shorter t$_{pr}$ can also lead to incomplete blow-down from the depressurizing column. Thus, the process pressure ratio might be lower, leading to poor device performance. On the other hand, larger t$_{pr}$ allows for sufficient pressure buildup in the system, thus enhancing the peak pressure values. It also allows for a complete discharge of the depressurizing column. Nonetheless, an optimal t$_{pr}$ value is desired as very large values would allow the non-uniform oxygen front to escape the adsorbent column, thereby directly affecting the purity. For the present system, as shown in Fig.~\ref{fig:Pres_curve}, the above competing scenarios lead to an optimality at t$_{pr}$ = 26 s, wherein one can expect balanced performance, stability, and good operational efficiency. Figure~\ref{fig:Pres_curve} also reveals an excellent match between the experimental and numerical oxygen purities at the outlet. The error bar in the experimental data includes the O$_2$ sensor's uncertainty error of $\pm$ 0.5 \% and the small fluctuations observed in the output purity.


\begin{figure}[!ht]
    \centering
     \includegraphics[width=12cm]{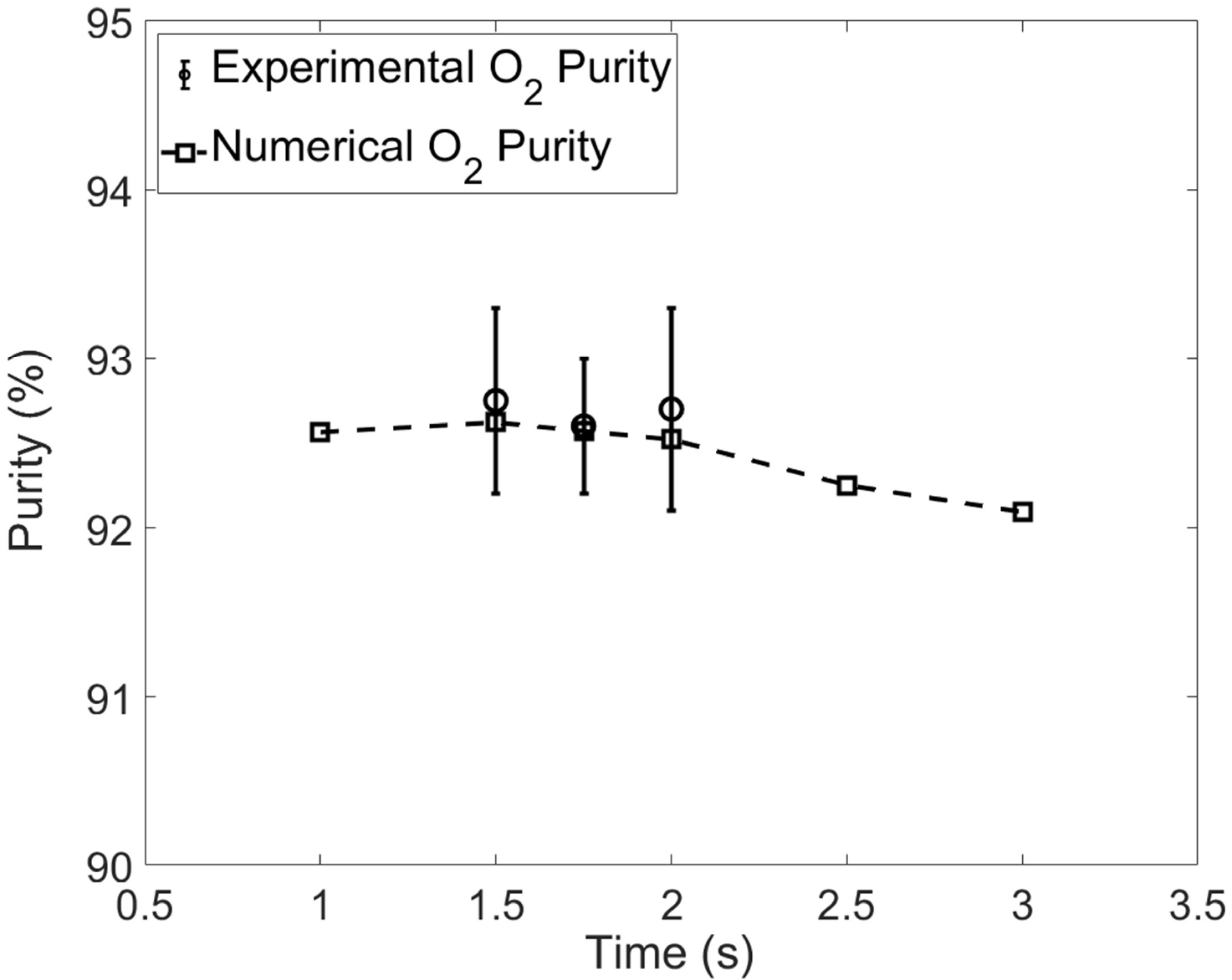}     
    \caption{Effect of purge time on purity}
    \label{fig:pruge_curve}
\end{figure}

\begin{figure}[!ht]
    \centering
     \includegraphics[width=12cm]{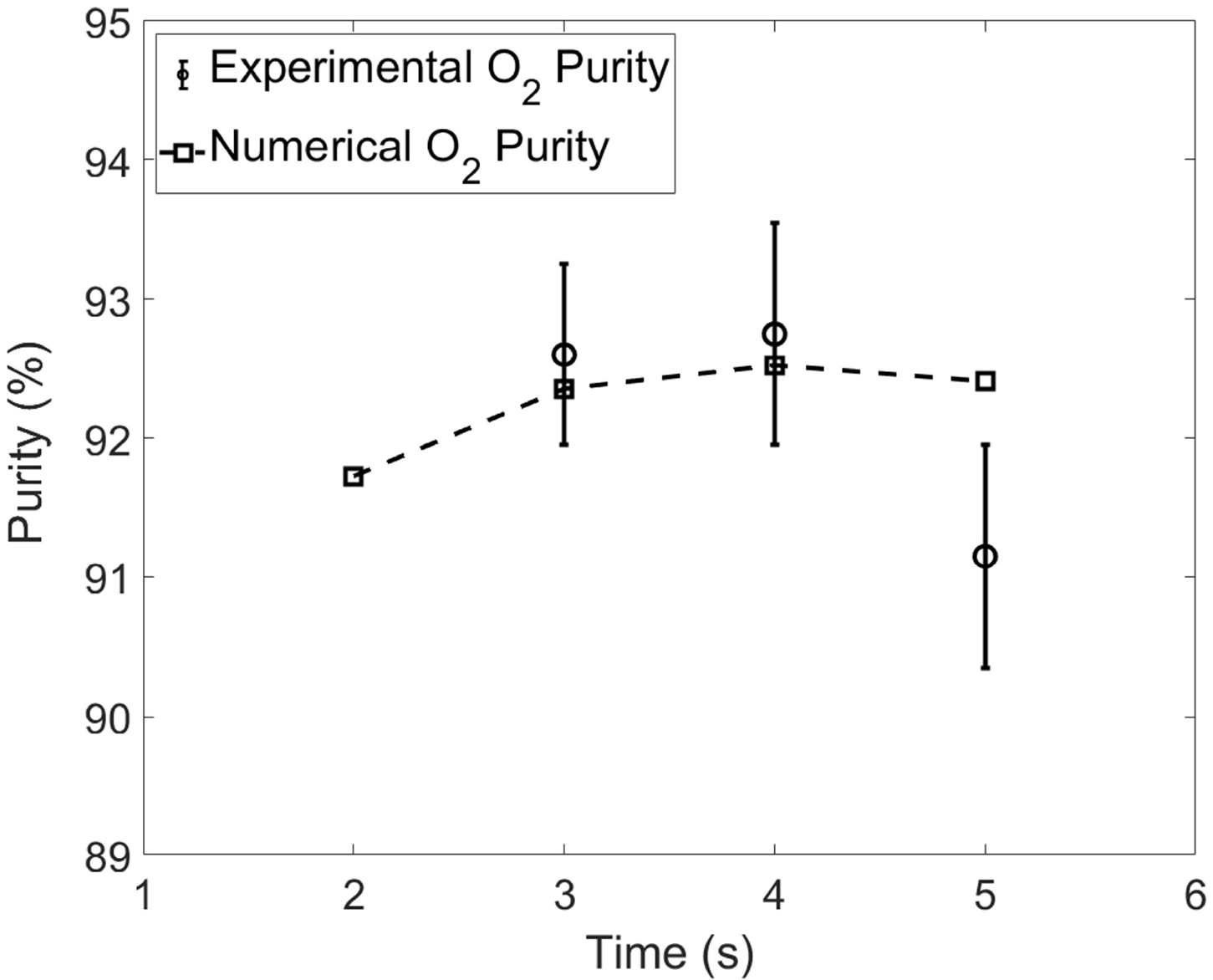}     
    \caption{Effect of equalization time on purity}
    \label{fig:eq_curve}
\end{figure}

\subsection {Effect of purge time}
Continuing with the process optimization, we now analyze the influence of changing t$_{pu}$ on the overall device performance. To explore this, we vary the purge time from 1 s to 3 s, with t$_{pr}$ being fixed at 26 s and t$_{eq}$ at 4 s. Similar to the pressurization case study, Fig.~\ref{fig:pruge_curve} illustrates a strong agreement between the experimental and numerical data. The results indicate that the t$_{pu}$ value between 1.5 s and 2 s provides an optimal balance between effective regeneration of the adsorbent and operational efficiency. Shorter purge times result in inadequate flushing. On the other hand, longer purge times, though allowing for a thorough regeneration of the adsorbent, result in wasting the precious end product.  

\subsection {Effect of equalization time}
Finally, we analyze the optimal time required for properly equalizing pressure between the columns. In this regard, t$_{eq}$ has been varied from 2 s to 5 s, with t$_{pr}$ fixed at 26 s and t$_{pu}$ at 2 s. The results plotted in Fig.~\ref{fig:eq_curve} indicate that an equalization time of 4 s provides an optimal balance between pressure equalization and operational efficiency. Shorter equalization times result in incomplete pressure equalization and, thus, suboptimal performance. In contrast, larger t$_{eq}$ values, while improving pressure equalization, cause increased fluctuations in the oxygen storage tank pressure, hence resulting in an unstable output flow rate.

\section {Conclusion}
The current work presents an efficient and simplified digital twin model for mimicking and optimizing commercial-type PSA plants. The model involves a meticulously configured 2D axisymmetric representation that emulates all the essential components of the PSA system, such as the adsorbent columns, solenoid valves, pressure regulators, and mesh filters, using suitable porous zone approximations. The adsorption process in the zeolite columns is estimated using the LDF model and is integrated into simulations via UDFs. With appropriate transient modifications to the boundary conditions for mimicking solenoidal valves, the model offers a comprehensive framework to understand the PSA system's dynamic behavior. It is important to note that the model does not have any parameters that need to be fitted for its closure. The capability of this digital twin has been established here by comparing its results with those of an in-house PSA pilot plant used for oxygen separation. Here, an excellent match of the pressure buildup characteristics and the output purity has been observed for different operating conditions. The model also reveals intricate details of the device's operation, providing invaluable insights for device optimization. For example, two suggestions for improving the in-house PSA plant are already evident from the detailed data of the digital twin. The first suggestion pertains to the avoidance of bottom-bottom equalization to prevent re-entry of N$_2$ in the adsorbent columns, while the second involves modification of valve operation to prevent backflow from the O$_2$ storage tank to the adsorbent columns. The model reveals its sensitivity through its precise response to changes in variables such as pressurization, purge, and equalization times, and its ability to identify the optimal values of these parameters, thereby striking a delicate balance between operational efficiency, energy consumption, and product purity. In summary, the digital twin, with its ability to precisely mirror the real-world performance of the PSA system, offers a potent tool for designing and improving gas separation technologies in various industrial sectors.


 \bibliography{acs-achemso}





\end{document}